# Methane on Mars: new insights into the sensitivity of CH$_4$ with the NOMAD/ExoMars spectrometer through its first in-flight calibration

Giuliano Liuzzi[a,b], Geronimo L. Villanueva[a], Michael J. Mumma[a], Michael D. Smith[a], Frank Daerden[c], Bojan Ristic[c], Ian Thomas[c], Ann Carine Vandaele[c], Manish R. Patel[d,e], José-Juan Lopez-Moreno[f], Giancarlo Bellucci[g], and the NOMAD team

[a] NASA Goddard Space Flight Center, 8800 Greenbelt Rd., Greenbelt, 20771 MD, USA
[b] Dep. of Physics, American University, 4400 Massachusetts Av., Washington, 20016 DC, USA
[c] Belgian Institute for Space Aeronomy, BIRA-IASB, Ringlaan 3, 1180 Brussels, Belgium
[d] School of Physical Sciences, The Open University, Milton Keynes, MK7 6AA, UK
[e] SSTD, STFC Rutherford Appleton Laboratory, Chilton, Oxfordshire OX11 0QX, UK
[f] Instituto de Astrofisica de Andalucia, IAA-CSIC, Glorieta de la Astronomia, 18008 Granada, Spain
[g] Istituto di Astrofisica e Planetologia Spaziali, IAPS-INAF, Via del Fosso del Cavaliere 100, 00133 Rome, Italy

Corresponding author: Giuliano Liuzzi, e-mail: giuliano.liuzzi@nasa.gov

## Abstract

The Nadir and Occultation for MArs Discovery instrument (NOMAD), onboard the ExoMars Trace Gas Orbiter (TGO) spacecraft was conceived to observe Mars in solar occultation, nadir, and limb geometries, and will be able to produce an outstanding amount of diverse data, mostly focused on properties of the atmosphere. The infrared channels of the instrument operate by combining an echelle grating spectrometer with an Acousto-Optical Tunable Filter (AOTF). Using in-flight data, we characterized the instrument performance and parameterized its calibration. In particular: an accurate frequency calibration was achieved, together with its variability due to thermal effects on the grating. The AOTF properties and transfer function were also quantified, and we developed and tested a realistic method to compute the spectral continuum transmitted through the coupled grating and AOTF system. The calibration results enabled unprecedented insights into the important problem of the sensitivity of NOMAD to methane abundances in the atmosphere. We also deeply characterized its performance under realistic conditions of varying aerosol abundances, diverse albedos and changing illumination conditions as foreseen over the nominal mission. The results show that, in low aerosol conditions, NOMAD single spectrum, 1$\sigma$ sensitivity to CH$_4$ is around 0.33 ppbv at 20 km of altitude when performing solar occultations, and better than 1 ppbv below 30 km. In dusty conditions, we show that the sensitivity drops to 0 below 10 km. In Nadir geometry, results demonstrate that NOMAD will be able to produce seasonal maps of CH$_4$ with a sensitivity around 5 ppbv over most of planet's surface with spatial integration over 5x5 degrees bins.

Results show also that such numbers can be improved by a factor of ~10 to ~30 by data binning. Overall, our results quantify NOMAD's capability to address the variable aspects of Martian climate.

# 1 Introduction

Our understanding of Mars and its evolution has grown tremendously in the last decades, thanks to unique data from a wide range of missions and observations. However, most instruments and missions have been dedicated to a comprehensive characterization and global mapping of its surface geology and landforms, leaving aside a detailed analysis of the atmospheric composition, especially in terms of trace gases. Of instruments investigating the Martian atmosphere, we mention the Thermal Emission Spectrometer, TES (Christensen et al., 2001) on board the Mars Global Surveyor (MGS), the Compact Reconnaissance Imaging Spectrometer for Mars, CRISM (Murchie et al., 2007) and the MARs Color Imager, both operated on board the Mars Reconnaissance Orbiter (MRO), and three instruments on Mars Express (MEX): PFS (Planetary Fourier Spectrometer, Formisano et al., 1997), SPICAM (Spectroscopy for Investigation of Characteristics of the Atmosphere of Mars (Bertaux et al., 2006)), and OMEGA (Observatoire pour la Minéralogie, l'Eau, les Glaces et l'Activité, (Bibring et al., 2006).

The most valuable scientific breakthroughs achieved with TES data involve a quite complete characterization of surface composition and spectral properties and related methods (Bandfield et al., 2000; Bandfield and Smith, 2003; Liuzzi et al., 2015; Smith, 2008) and an unprecedented understanding of the vertical structure of atmosphere (temperature and aerosols distribution) and its global dynamics (Conrath et al., 2000; Smith, 2004; Wolff, 2003). CRISM data have revealed further insights into the characterization both of surface landforms and mineralogy (Byrne et al., 2009; Pelkey et al., 2007), and about the climatology of aerosols and water vapor in the atmosphere. As far as MEX instruments are concerned, instead, although their major focus was the detection of a variety of trace gases, only detections of very few have been claimed (among them $CH_4$ column abundance and $O_2(a^1\Delta_g)$ in the upper atmosphere, see e.g. Montmessin et al., 2017; Perrier et al., 2006). Hence, at present, no space mission has been really tailored to the extensive detection and quantification of trace gases, apart from the very specific issue of methane detection in the Martian atmosphere, which has generated heated debate in the scientific community (see e.g. Fonti et al., 2015; Formisano et al., 2004; Krasnopolsky et al., 2004; Mumma et al., 2009).

To date, the only observations dedicated to trace gases measurements and mapping measures are ground-based (e.g. Khayat et al., 2017; Mumma et al., 2009; Villanueva et al., 2013). The Earth-facing hemisphere can be mapped with long-slit spectrometers, providing a snapshot in time/season of the targeted trace gas (and $CO_2$), examples include $CH_4$, $H_2O$ and HDO (Villanueva et al. 2015), and $O_3$ through (its photolysis product) $O_2(a^1\Delta_g)$ (Novak et al., 2002) Recent measurement campaigns have been conducted using the iSHELL spectrograph at the NASA InfraRed Telescope Facility observatory in Hawaii. Despite the very high quality of iSHELL (with a resolving power ~70000), ground-based observations cannot provide a truly comprehensive, global and continuous temporal picture of the abundances, global distributions and possible local release of individual trace gases on Mars. Ground-based observations are reserved to restricted time periods, since they require enough spectral Doppler shift between Mars and Earth to avoid severe blanketing of the Mars gaseous signatures by terrestrial counterparts, and the requested observing interval is not always awarded. Furthermore, the mapping capabilities are very limited in Mars polar regions, and in general are usually limited to a single hemisphere of the planet. Despite this, some Earth based observations have been useful to clearly identify trace gases and to map their spatial distributions, on a global scale; important examples include $CH_4$, $H_2O$ and HDO, and $O_2(a^1\Delta_g)$. The derived parameters (e.g., D/H in Mars water) provide key evidence for understanding the past evolution of the planet (e.g. Villanueva et al., 2015), and its present activity, and can be related to disequilibria in the chemical composition of the atmosphere, as evidenced in some studies (e.g. Allen et al., 2006). In particular, the existing observations of methane are not complete. Methane detections by the Tunable Laser Spectrometer (Mahaffy et al., 2012) on board Curiosity (NASA's Mars rover), reveal pulsed release to levels (~7 ppbv) consistent with plumes detected in earlier ground-based results, but also a quite variable methane concentration at background levels (<1 ppbv) that could not be sensed by previous remote detections (Webster et al., 2015, 2018). Such variations could be a possible clue for biological processes. Moreover, the present knowledge of the Martian atmosphere does not fully address the question of geophysical activity on Mars, through the analysis of volcanic outgassing (Khayat et al., 2017).

The observation of trace gases in the Martian atmosphere is valuable also for achieving a better quantitative description and validation of their spectroscopic properties. Given its importance for tracing the evolution of the planet, this work has been done for select bands of isotopic $CO_2$ and for water and its isotopologues (e. g., Villanueva et al., 2008, Villanueva et al., 2012), but for many other trace gases a refinement of their spectral properties remains a work in progress.

The NOMAD spectrometer was selected as one of the four instruments for the ExoMars Trace Gas Orbiter mission to address these issues, and it investigates the atmosphere and the Martian surface in the UV, visible and IR wavelength ranges. NOMAD operates in 3 channels: an ultraviolet/visible channel (UVIS) which covers the spectral interval 0.2 – 0.65 µm and can work both in nadir and solar occultation geometry; and two infrared channels – one dedicated to solar occultation (SO) measurements in the infrared (nominal range 2.3 – 4.3 µm), and a second infrared channel observing in the limb and nadir geometries (LNO, 2.3 – 3.8 µm). Since the instrument is devoted to study atmospheric composition, it offers spectral resolving power ($\lambda/\delta\lambda$ ~20,000) far higher than any instrument previously operated in Martian orbit.

This paper is organized in two distinct parts: in the first, we focus on the spectral calibration of the two infrared channels of NOMAD, the SO and LNO. Their design is entirely inherited from the structure of infrared solar occultation channel of the Solar Occultation in the IR instrument (SOIR) on board of Venus Express (Nevejans et al., 2006). The concept of the instrument is based on the combination of a diffraction grating and an Acousto-Optical Tunable Filter (AOTF) which selects the principal diffraction order to be observed. We use the first in-flight data acquired by the instrument during the first Mars Capture Orbit phase to achieve a complete characterization of both the grating and the AOTF properties, as they were then operated in the same conditions in which the instrument will observe the planet. These elements are then combined in a mathematical model that computes NOMAD spectra of gases and their spectral continua, taking account of the flux contribution coming from adjacent diffraction orders (the AOTF admits throughput of orders adjacent to the targeted one, albeit with lower transmittance, and this algorithm permits identification, separation and modeling of the various components).

These results are then used in the second part, where we perform an analysis of the instrument capabilities in detecting $CH_4$ in solar occultation geometry, at different levels of aerosol opacity, and in Nadir geometry for different illumination conditions and surface albedos, accounting also for aerosol opacity. These problems are recently assessed in Vandaele et al., 2018 for other gases together with $CH_4$ and its different isotopes. The results we present here benefit from the exhaustive instrumental characterization described in the first part, which relies on in-flight data. This is the first work which makes use of NOMAD in-flight data.

In addition, although NOMAD IR channels have a lot in common with SOIR design, they also enable the possibility to observe the planet in nadir geometry.

In Section 2 we outline the main characteristics of NOMAD IR channels, with some focus on the optics and the AOTF operating principles, and a short illustration of the structure of the data that will be used in this work. Section 3 is dedicated to the description of calibration procedures used to determine grating properties, instrument spectral resolution, and AOTF bandpass shape and spectral properties. These results are used to build a model for the continuum observed in NOMAD data, whose details are in Section 4. Section 5 addresses the actual sensitivity of NOMAD to atmospheric methane, computed using full radiative transfer calculations of NOMAD spectra in Solar Occultation geometry at different heights and aerosol loads. Although studied in other cases (Robert et al., 2016; Vandaele et al., 2018), this new analysis now considers more realistic information by using the latest in-flight calibration of the channels, and refines the results presented in the other studies. This serves as a proxy to depict the seasonal trend of $CH_4$ detection limits, according to planetary climatology. Section 6 presents a similar analysis, on the spatial domain, for LNO nadir observations.

## 2  Instrument and data description

### 2.1  Instrument

The infrared channels (SO and LNO) of the NOMAD instrument have been described in great detail by Neefs et al., 2015 and Thomas et al., 2016, and only a short summary will be reported here. The concept is that of a compact echelle grating spectrometer, which is arranged in a Littrow configuration and is combined with a $TeO_2$ AOTF for pre-selection of the diffraction order to be observed. This is done by addressing the AOTF filter with a radio frequency (RF) generator, which modifies the diffraction properties (namely, the refractive index) of the AOTF as the input RF varies. The AOTF is a narrow bandpass filter and the input RF selects the central frequency at which the AOTF transfer function peaks, and consequently the diffraction order that enters the spectrometer slit and later falls upon the detector.

The NOMAD detector is made up of a grid of 320 columns (the spectral direction) and 256 rows (the spatial direction), and for a given the AOTF setting the instrument ideally projects a single diffraction order on the detector. In this respect, the choice of the AOTF bandwidth and response is critical; in principle, it should be a square wave equal to or narrower than the Free Spectral Range of the echelle spectrometer, which for NOMAD is equal to 22.55 cm$^{-1}$ (Neefs et al., 2015). In practice, the AOTF transfer function is likely characterized by 'wings' to shorter and longer wavelengths of the main lobe (Neefs et al., 2015), which introduce a spurious flux coming from wavelengths beyond the Free Spectral Range. The grating efficiency is much smaller at these frequencies, but some photons are transmitted by the AOTF and are detected. A qualitative description of this effect is depicted in Figure 1, and a complete treatment of it is left for the forthcoming sections.

The observation geometry of NOMAD is illustrated in Figure 2. The SO channel operates with the instrument pointed towards the Sun, in order to observe the solar radiation as it is successively attenuated by the Martian atmosphere at different altitudes and thus to investigate the vertical structure of the atmosphere. Observations can be performed both in ingress (at the sunset with respect to the point of view of the spacecraft) or in egress (sunrise). Since ExoMars TGO is in a polar orbit (12 orbits per Sol), the NOMAD instrument can ideally perform two SO observations per orbit, for a total of 24 occultations per Sol. The theoretical spectral interval of operation in SO configuration is 2.3 – 4.3 µm (2325 – 4350 cm$^{-1}$, corresponding to the diffraction orders from 96 to 225), with a theoretical resolving power of ~20000, and the input RF to AOTF varying between ~12300 to ~31100 kHz.

The LNO operation mode is instead conceived to observe the planet in a nadir or limb geometry, switching between the two modes by rotating the spacecraft. The LNO channel is able to perform a complete mapping of the planetary surface every 30 sols. Observing the surface instead of the Sun implies that the signal to be detected is much weaker than that observed in SO mode, requiring some modifications in the instrumental configuration. In particular, the LNO channel is characterized by a larger entrance slit, and most of the optical elements are larger. Besides this, a much longer accumulation time (15 s against ~1 s in SO mode) is needed to raise the signal to noise ratio up to ~100 in nadir geometry. The reduction of the signal, and the challenges to enhance the Signal-to-Noise Ratio (SNR) value for LNO has some counterparts in the effective spectral interval spanned by the instrument (reduced to 2.3 – 3.8 µm or 2630 – 4350 $cm^{-1}$, in diffraction orders from 108 to 220), with a theoretical resolving power reduced to ~10000, and the input RF to AOTF ranging between ~14200 and ~32100 kHz.

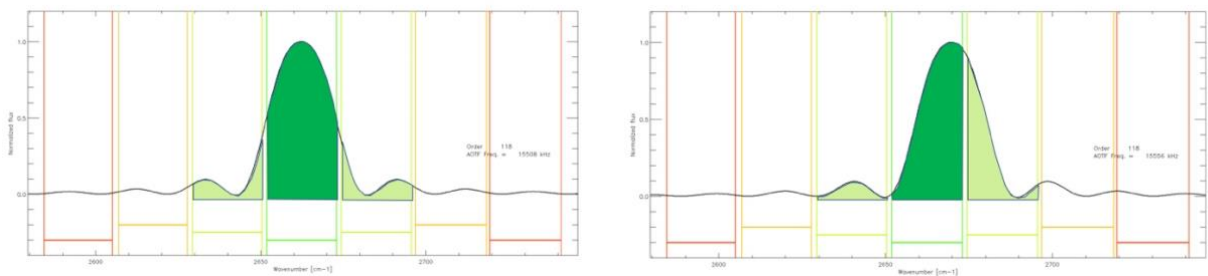

*Figure 1 - Example of variation of order overlapping and variation of contribution from side orders. In the second case (right panel) the AOTF transfer function is shifted with respect to the order center, yielding an invasive contamination by the shortwave nearby order. The flux from nearby orders is minimized if the AOTF input radio frequency is chosen in order to center the AOTF transfer function with respect to the observed order (left panel).*

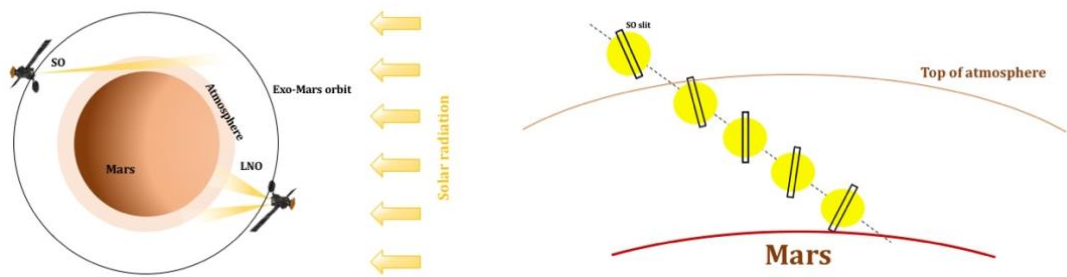

*Figure 2 – Left panel: Overview of the NOMAD SO and LNO viewing geometry. For the LNO, the scheme reports both the pictures of nadir and limb geometry. Right panel: scheme of a typical SO observation: the slit rotates with respect to the perpendicular to the surface as the altitude of Sun and Line of Sight (LOS) changes. The slit is typically larger than the Sun disk. The effect of slit rotation is exaggerated for illustrative purposes.*

The characteristics of NOMAD have imposed constraints on the routine science operations, started in April 2018. Thanks to the flexibility in switching between different diffraction orders (granted by AOTF filtering), measurements can consist of rapidly repeated cycles of 1 to 6 key diffraction orders whose spectral ranges include absorption by gases/organics of interest. Instead, the limitation in the amount of data to be transferred, together with the need to raise the SNR, limits the number of rows (effective or binned) of the detector output to 24. This is the same strategy previously adopted by SOIR.

Independently of the way in which spatial binning is performed in calibrated data, NOMAD field of view varies according to the type of observation and channel. In the case of an LNO nadir observation, for a typical total integration time of 15 s, the total footprint on the surface is never larger than 1° x 0.3°, or equivalently 51 x 17.5 km$^2$. For solar occultation measurements with the SO channel, the 24 rows of the detector typically cover an interval of ~7.5 km along the tangent height path. Hence, a single detector row will typically sample a vertical distance of ~500 m.

Finally, as far as calibration is concerned, it is necessary to note that the SO and LNO channels work with two different AOTFs and slightly different optical configurations. Hence, data acquired in SO and LNO configurations will be treated separately, as coming from two different instruments, though elaborated with the same procedures.

The reader is referred to Neefs et al., 2015 for a more extensive description of the instrument.

## 2.2 Datasets

The observations for the calibration procedures presented here have been acquired during the First Mars Capture Orbit, between 22$^{nd}$ and 27$^{th}$ of November 2016. All the observations considered are calibration measurements, in which the instrument points directly to the Sun, both in SO and LNO configurations; a small number of observations were performed in nadir mode to measure the atmospheric transmittance, but they are not addressed in this calibration phase. In these calibration measurements, integration times per single acquisition were always between 1 and 4 milliseconds (ms) to prevent detector saturation, and a number of measurements between 64 and 86 were accumulated. According to the way in which AOTF input RF is varied, measurements can be of two different types:

- **Miniscan**: The NOMAD SO and LNO channels perform a sweep over a fraction of their spectral range whilst pointing towards the Sun. To do this, the AOTF input RF is slowly varied (at steps of 2 kHz), in order to observe the behavior of the observed signal across different orders and in the transition between them. These are the observations that have been actually employed for the calibration procedures described in this paper.

- **Full scan**: During a 'Full scan' observation, the spacecraft points the NOMAD nadir boresights to nadir (in LNO) or the Sun through the atmosphere (SO), and the AOTF input RF is varied by large steps (~160 kHz) to perform a sweep over its complete spectral range, one diffraction order at a time. These measurements have been employed only to quantify thermal effects on wavenumber calibration, together with miniscans, and used as test datasets for the solutions obtained.

Nadir, limb, and solar occultation routine measurements (i.e. not used for calibration purposes) are made using the principles of Full Scan, hence measuring sequentially one or more diffraction orders at specific AOTF input RF (from now on, "AOTF frequencies"). Examples of how the observed signal behaves in Miniscan or Full Scan modes are given in Figure 3.

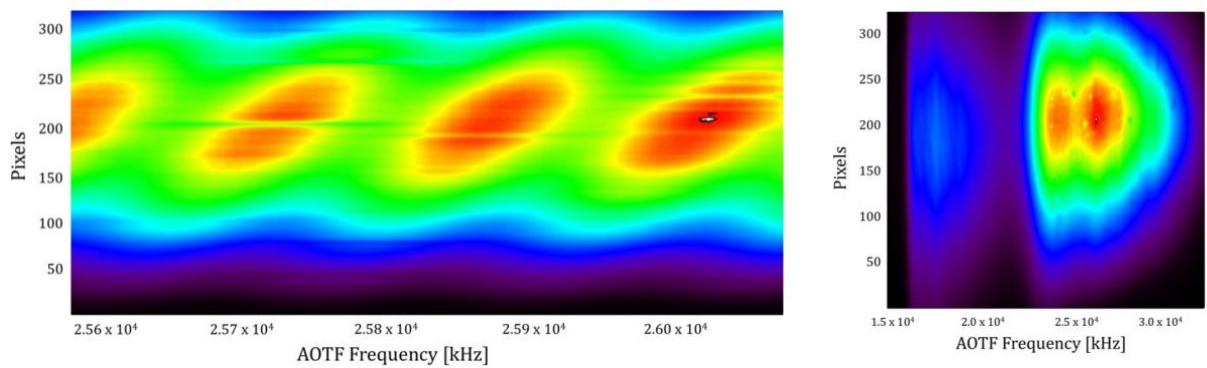

*Figure 3 – Left panel: Example of LNO miniscan data (diffraction orders 177-180). The intensity of the signal is represented by colors (blue to red as signal increases). Solar lines are evident as horizontal stripes. The transition between orders as AOTF frequency increases corresponds to the intervals, along X axis, between consecutive peaks. Right panel: example of LNO full scan data. In this case, every column of the image represents a single spectrum, and a particular diffraction order. As before, the absorption due to solar lines is visible in the image (now as 'spots'); the variation of signal intensity with AOTF frequency across the entire spectral interval sounded by NOMAD (LNO in this case) is also evident.*

## 3 NOMAD full spectral and AOTF calibration

The calibration of NOMAD spectra, as said, involves both the AOTF and grating characteristics. The list of calibrations we perform are the following:

- Conversion of detector pixel number and diffraction order to wavenumber, namely the spectral calibration of the instrument;
- Determination of the spectral resolution for SO and LNO, hence their effective resolving power;
- Quantification of thermal effects on the spectral calibration, namely the shift of frequency across the detector due to thermal-induced mechanical stress;
- Measurement of the "tuning relation", namely the relation between the radio frequency applied to AOTF and the wavenumber at which the AOTF transfer function peaks;
- Measurement of the shape of the AOTF transfer function;

These elements will be used, together with the grating blaze function (Thomas et al., 2016), to build a model for the spectral continuum as seen by NOMAD (see Section 4), and applied to NOMAD data for their normalization and simulation. An *a priori* knowledge of the spectral continuum is of particular importance for a correct normalization of NOMAD nadir data, for which no "background" observation (i.e. free of atmospheric absorption) is possible, by contrast with occultation geometry.

## 3.1 Spectral calibration and resolving power estimation

The principle of spectral calibration is based on the position of well-defined solar lines in the calibration spectra, that can be used to match a line's frequency with the pixel number in which it falls. The aim of this calibration is to obtain an analytic relationship between the pixel number of the detector, the diffraction order, and the corresponding wavenumber. In this process, the order number is not an independent variable, since it is fixed by AOTF input frequency $A$. Together with the geometric and optical properties of the grating and the optics of the instrument, this is the only independent variable in this calibration scheme, together with the pixel number. As a by-product of this process, an estimation of the resolving power of the instrument is also retrieved.

The relation between pixel number $p$ (from 0 to 319), wavenumber $v$ and order number $m$ is modeled by a second-order polynomial of the form:

$$\frac{v}{m} = F_0 + F_1 p + F_2 p^2 \tag{1}$$

while the tuning relation between the AOTF frequency $A$ and the wavenumber $v$ at which AOTF bandpass is centered is

$$v = G_0 + G_1 A + G_2 A^2 \tag{2}$$

To perform the spectral calibration, we used the spectra acquired in Miniscan mode, and averaged all those whose AOTF frequency falls within the same diffraction order (inferred from the combination of Eq. (1) and (2)).

Then, within an iterative process, the observed spectrum $S_0$ was shifted in frequency (of an amount $\Delta v$) by regular steps (equal to the spectral sampling, i.e. one pixel), in order to seek the best correlation between the whole spectrum $S_0$ and a synthetic solar spectrum $S_\odot$ computed on the same spectral interval (Hase et al., 2010), and between the part of the observed spectrum $S_{0L}$ that contains the brightest solar line and the synthetic one $S_{\odot L}$. Hence, the two highest correlations, namely $C_0$ and $C_L$, are computed:

$$C_0 = max(corr(\boldsymbol{S_0}, \boldsymbol{S_\odot})) \tag{3}$$

$$C_L = max\left(corr(\boldsymbol{S_{0L}}, \boldsymbol{S_{\odot L}})\right) \tag{4}$$

and the corresponding displacement $\Delta v$ with respect to the pre-computed calibration is the minimum of the two displacements corresponding to the two correlation values.

In doing the selection of the solar lines for fitting Eq. (1), we discard all those lines that are superimposed to the ghost images of other solar lines observed in ±2 nearby diffraction orders, to reduce the number of outliers in the fit. For the same reason, the analysis discards all the solar lines for which the retrieved center is not within 0.5 cm$^{-1}$ of the first estimate; also, we include only lines whose intensity is above a defined threshold (i.e., total integrated line opacity, parameter A as defined by Hase et al., (2006), is above 0.4).

The double correlation expressed in Eqs. (3) and (4) makes the procedure less affected by a possible periodicity in the solar lines, or in cases in which two solar features of similar intensities occur within a few pixels. The output of this first step is a new spectral grid, denoted with $v + \Delta v$.

Once this first step is performed, the actual wavenumber calibration is done by locating the brightest solar lines; the pixel at which a line is located is denoted as $p_0$. Each single line is then fitted using a Gaussian model, which defines the center of the line (in "absolute wavenumber" units, $\tilde{v} = (v + \Delta v)/m$), and its Full Width at Half Maximum (*FWHM*). An estimate for the three coefficients in Eq. (1) is then provided by fitting the relation between the center of the lines (values of $p_0$) and the central wavenumbers of the same lines divided by the corresponding diffraction orders, $v_0/m$. $v_0$ values are derived fitting the lines with Gaussians, along with the corresponding errors. For each of the line considered, the resolving power is computed as:

$$RP = \frac{v_0}{FWHM} \qquad (5)$$

To retrieve calibration coefficients, we have used all the Miniscan acquisitions available between the 22$^{nd}$ and 27$^{th}$ of November 2016, to cover almost continuously the diffraction orders observed by NOMAD (99-211 for SO and 108-210 for LNO), for a total of ~70 solar lines. Results of the fitting are represented in Figure 4 and Figure 5, where each point corresponds to the position (in terms of pixel and $\tilde{v}$) of a single solar line, while its color indicates the diffraction order in which that solar line is observed. The bottom panel reports the residuals with respect to the retrieved wavenumber calibration. Residuals indicate that the calibration error for the position of lines is always lower than 0.3 cm$^{-1}$, and the average residual is ~0.07 cm$^{-1}$ for SO and ~0.05 cm$^{-1}$ for LNO, and both correspond to less than one pixel.

Retrieved coefficients for the spectral calibration in Eq. (1) are reported in Table 1. The numbers for LNO channel are coincident with those based on laboratory calibration data, observing the absorption lines due to a gas cell (Thomas et al., in preparation).

Equation (1) is useful to verify the consistency of the properties of the instrument derived during the ground calibration. In particular, the Free Spectral Range can be derived directly by the equation itself, and it is equivalent to the second member of Eq. (1) computed at pixel 160 (center of the order). The result is 22.563 cm$^{-1}$ for SO and 22.567 cm$^{-1}$ for LNO, which are very close to the values reported in Neefs et al., 2015 rescaled with the instrument temperature.

The second output of the procedure is the resolving power of the instrument, retrieved from the parameters derived by fitting the individual solar lines, performed with a least-squares approach. In this process, it is assumed that the line profile is independent of grating temperature, order, and pixel number. Lines are divided by the local continuum, and fitted with a Gaussian profile, without the addition of any continuum. The opportunity to choose other profiles (e.g., sinc) to fit the solar lines has been investigated, but results showed that a Gaussian profile can better reproduce the actual observed line shape, given the absence of side lobes in the solar lines; also, the core of the line is nicely reproduced by a Gaussian profile, rather than a sinc. This is consistent with the findings from SOIR (Vandaele et al., 2013). Further investigations on this issue by employing laboratory / ground data may prove more conclusive on the detailed line-shape response of the spectrometer.

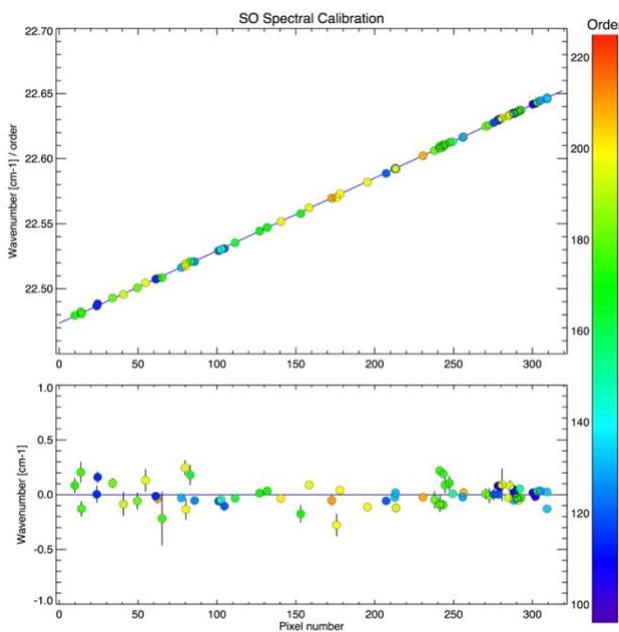

*Figure 4 - Results for all SO miniscans of 22nd, 23rd, 25th and 27th November 2016. Bottom panel reports residuals to the best fit.*

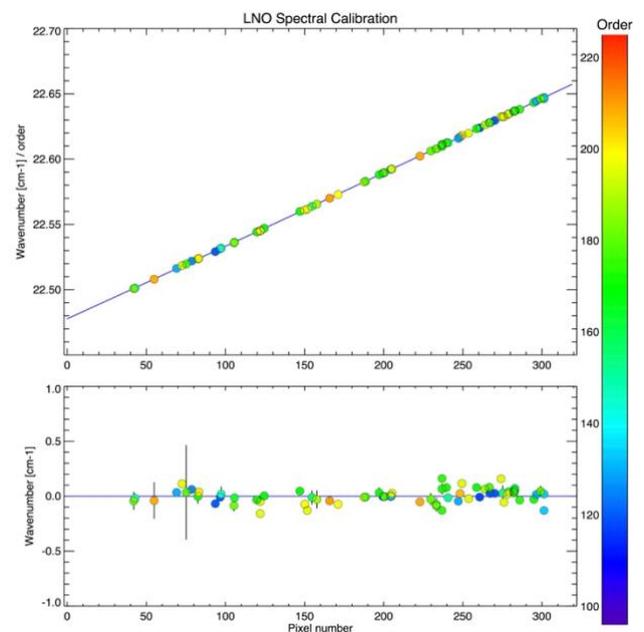

*Figure 5 - Results for all LNO miniscans of 22nd, 23rd, 25th and 27th November 2016. Bottom panel reports residuals to the best fit.*

| INSTRUMENT | DATE | $F_0$ | $F_1$ | $F_2$ |
|---|---|---|---|---|
| **SO** | 22$^{nd}$-27$^{th}$ Nov. 2016 | 22.473422 | 5.559526·10$^{-4}$ | 1.751279·10$^{-8}$ |
| **LNO** | 22$^{nd}$-27$^{th}$ Nov. 2016 | 22.478113 | 5.508335·10$^{-4}$ | 3.774791·10$^{-8}$ |

*Table 1 - Retrieved coefficients for the spectral calibration.*

Two examples of fits are provided in Figure 6, panels a) and b). Panels c) and d), instead, report the computed *FWHM* of different solar lines measured throughout the entire spectral interval covered by LNO (panel C) and SO (panel D). For SO, it has been found that the resolving power (see Eq. (5)) is ~19000, while for LNO it is ~11000. This is consistent with the difference in slit width, which is twice larger for LNO mode than for SO, hence for SO twice the resolving power than for LNO is expected. Both values are in good agreement with those found during ground calibration (Neefs et al., 2015). The linear relationship between frequency and *FWHM* of the lines, also, demonstrates that the resolving power is constant through the whole the spectral interval covered by NOMAD, within the precision of the calibration method.

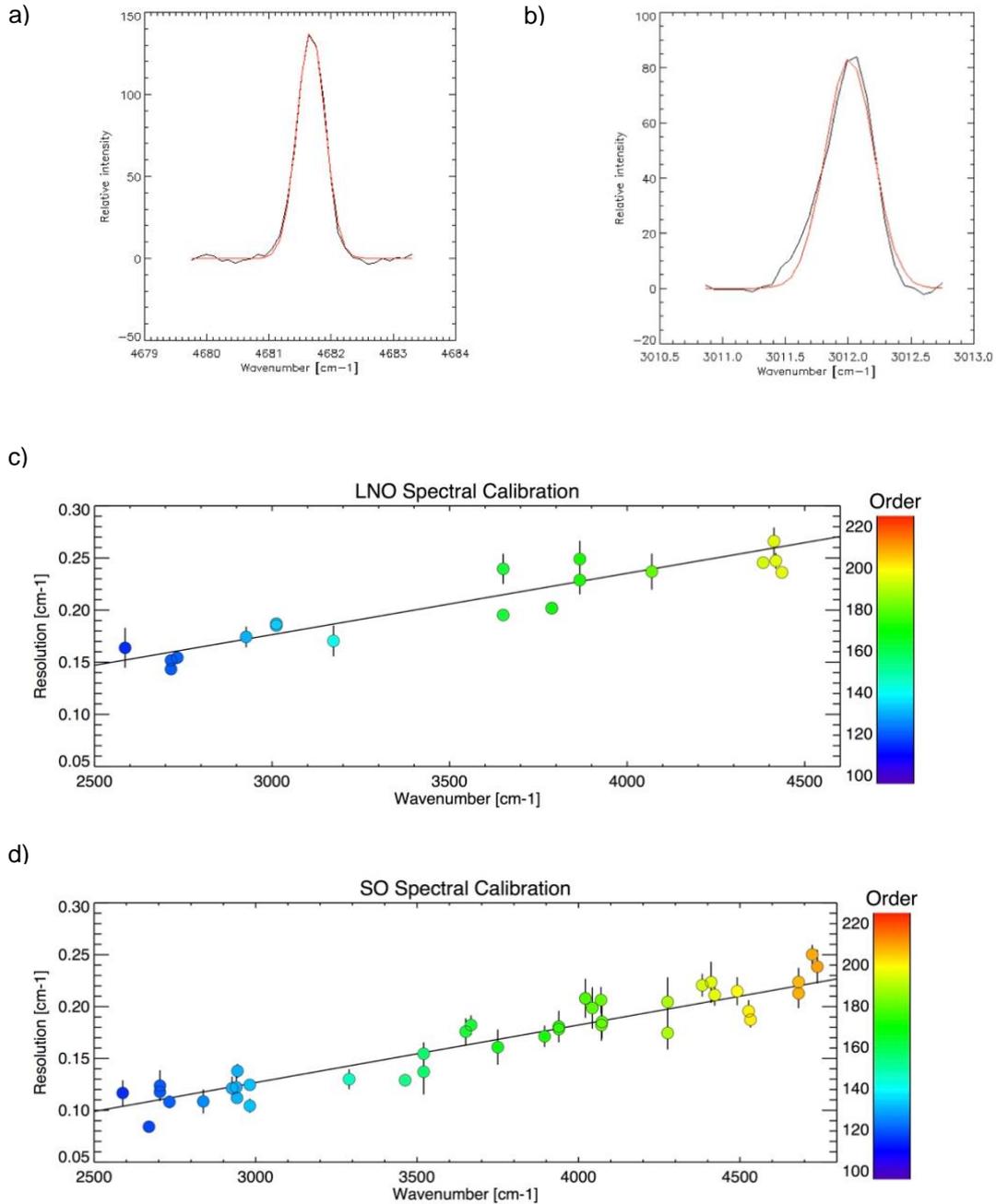

*Figure 6 – Panels a) and b): Examples of solar lines from LNO (left) and SO (right) miniscan observations. The corresponding Gaussian fit is in red. Panels c) and d): FWHM of the single solar lines as a function of the wavenumber. The linear relation between the two quantities reflects that the resolving power is constant (within the precision of our retrievals) throughout all the spectral interval covered by the instrument.*

## 3.2 Thermal effects on wavenumber calibration

The available observations have highlighted that the instrumental temperature is subject to variations of the order of ~10 °C from observation to observation. The dynamic ranges and absolute values of temperatures are generally different between SO and LNO configurations. Temperature fluctuations are responsible for mechanical strain in the grating support, which causes a displacement of the observed frequencies with respect to the frequency calibrations. This was observed in both laboratory measurements on gas cells and in the flight data described in the previous Section. The correction of these effects is included in the calibration scheme developed in this work, and is performed using a mechanism analogous to that implemented in the first part of frequency calibration. We use all the observations available for a certain date (both miniscans and full scans).

For each individual spectrum in each file, a first estimate for frequencies vs. pixel number and order is computed, based on the wavenumber calibration already performed. Then, through the same correlation scheme (Eq. (3) and (4)) implemented for spectral calibration, for each spectrum a displacement is computed with respect to the initial grid. As a last step, a polynomial fit is performed to correlate the displacements $\Delta p$ (in pixel unit), and the temperature $T$ (in °C) of the instrument, associated with each acquisition.

To cover the largest interval of temperatures and frequencies, a single fit has been performed, using all the available data. Results for SO and LNO are reported in Figure 7. A total of ~2500 spectra for LNO and ~3100 for SO have been used to retrieve the temperature shift.

It is clear that the displacement with respect to the wavenumber calibration can reach (absolute) values up to 8 pixels for SO (corresponding to ~1 cm$^{-1}$ for the higher diffraction orders), and up to 6 pixels for LNO. Despite this difference, both for SO and LNO we have retrieved a shift of ~0.75 pixels per degree Celsius. This is consistent with the fact that the materials and architectures of SO and LNO gratings and channels are quite similar, and in agreement with the value of 0.90 pixels per degree Celsius, which could be derived by the groove spacing of the grating and the geometric properties of the optical scheme of SO and LNO instruments (Neefs et al., 2015).

A very close result has been obtained utilizing lab data (Thomas et al., in preparation), which for LNO cover a wider temperature range (from -20 °C to +15 °C); thermal effects will be further investigated as new in-flight calibration data are acquired during current science operations.

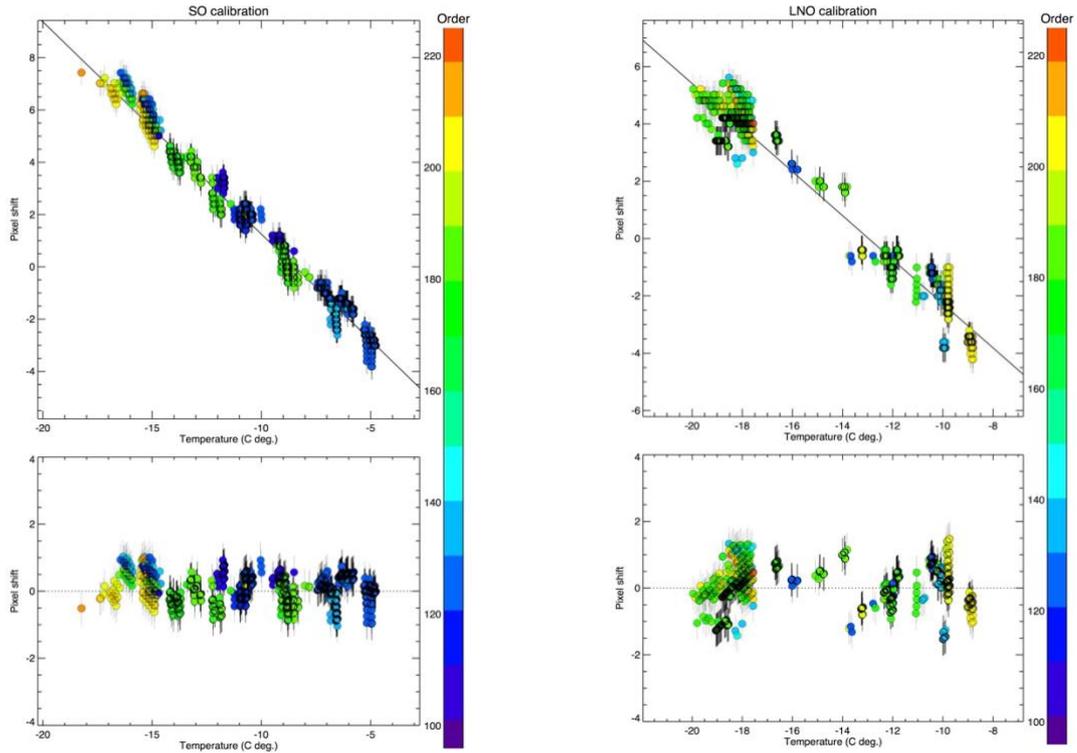

*Figure 7 - Fits between the frequency shift (in unit of pixels, Y-axis) and the temperature of the instrument (Degrees C, X-axis). Each point corresponds to a single spectrum. Results for SO are on the left, for LNO on the right. The color of each point corresponds to the diffraction order of the spectrum (see color scales). Bottom panels report the residuals to the best fit solution, which appear to be always within 1 pixel for SO and 1.5 pixels for LNO.*

It has finally to be stressed that, based on the fit residuals in Figure 7, the overall maximum calibration error is limited to one pixel (maximum 0.12 cm$^{-1}$ for the higher orders) for SO and 1.5 pixel for LNO (maximum 0.18 cm$^{-1}$). The derived frequency solutions for SO and LNO have already been tested on the first science data, and are generally accurate within these limits at every temperature.

### 3.3 AOTF tuning relation and AOTF transfer function shape

At a fixed radio frequency, $A$, the AOTF selects only a narrow wavelength band. The central wavelength $\lambda$ of a given band is determined by the momentum matching condition and can be expressed by the following equation, which is the original form of the AOTF tuning relation (Chang, 1981; Chang and Katzka, 1981; Mahieux et al., 2008):

$$A = \frac{w \Delta n}{\lambda}(sin^4\theta_i + sin^2 2\theta_i)^{1/2} \qquad (6)$$

where $\theta_i$ is the incident angle of the wave, $w$ is the acoustic velocity, and $\Delta n$ is the birefringence of the crystal. Using this, and measuring the optical properties of the AOTF, a theoretical tuning relation can be obtained. Based on Eq. (6), one could expect that the AOTF transfer function is subject to changes with wavenumber (pixel and order, Mahieux et al. (2009)), hence lab calibration and in–flight characterization are necessary to verify whether or not this happens.

Adopting our strategy for the grating equation, this calibration exploits the data from the miniscans, which span an AOTF frequency interval of ~500 kHz at steps of 2 kHz, fine enough to capture every salient characteristic of shape in the AOTF transfer function. The basic idea is to track the depth of selected solar lines as AOTF frequency slightly varies throughout the orders covered by a miniscan (see Mahieux et al. (2008) for a description of the method). This enables a description both of the shape of the AOTF transfer function as a function of wavenumber, and of the AOTF frequency corresponding to the maximum sensitivity to the wavenumber at which the solar line is located.

As before, spectra are classified by order, using the approach explained in Section 3.1. The calibration of AOTF properties benefits from results obtained for the spectral calibration, so the spectral grid is calculated, for each order, using the coefficients $F_0$, $F_1$ and $F_2$ already derived, and some *a priori* values for $G_0$, $G_1$ and $G_2$ (from ground calibrations). For each order, all solar lines whose intensity is above a defined threshold are selected. Then, in order to perform homogeneous calculations independently of the order, the position of each selected solar line is expressed in terms of the "absolute wavenumber" $\tilde{v}$.

Issues can arise in cases where there are two or more solar lines in the same diffraction order, or ghost features of the nearby ±2 orders, that fall within the same pixels, or too close one to each other. This occurrence has to be avoided especially when these solar lines fall in different diffraction orders, because in this case the line depth as a function of AOTF frequency will exhibit as many peaks as the number of superimposed solar lines of different orders, preventing assignment of the correct correspondence between AOTF frequency and line position.

To avoid this circumstance, the selected lines are sorted by $\tilde{v}$ values, and those closer than the equivalent – in wavenumbers – of 5 pixels, are discarded.

The next step is, for each of the remaining $k = 1, ..., N$ lines, to calculate the line depth as a function of the AOTF frequency $A$, with the double aim stated at the beginning of the Section. To do so, the algorithm extracts the 30 pixels around the center of the line, and uses the pixels free of any absorption line to compute a local continuum $C_k$ around the $k$-th line, as a simple 3rd order polynomial, which has been proven to be stable with respect to the variability of the spectral slope. Then, the continuum is used to compute the integral $L_{d,k}(A)$ of line absorption, for each AOTF frequency $A$ in the miniscan:

$$L_{d,k}(A) = \frac{1}{2\Delta v + 1} \sum_{j=v_{0,k}-\Delta v}^{v_{0,k}+\Delta v} |C_k(j,A) - S(j,A)|, \qquad k = 1, ..., N \qquad (7)$$

where $\Delta v$ is the number of pixels around the line center used to compute the integral. In order to limit possible contribution due to flux coming from nearby orders, typically $\Delta v$ is set to 1. Each $L_{d,k}(A)$ function corresponds to the variation of depth of a single solar line as $A$ varies, hence it is representative of the AOTF transfer function, which can be fitted by a suitable model. The fit will provide both the parameters describing the AOTF transfer function shape, and the AOTF frequency at which the line depth has its maximum, which can be correlated to the frequency of the line, to retrieve the parameters in Eq. (2). The model by which each $L_{d,k}(A)$ function is fitted is much simpler than that used for SOIR (Mahieux et al., 2009), which was structured as the sum of five sinc-squared terms plus a polynomial continuum. Though the continuum has been kept also in this case to mitigate the effects of possible inconsistencies in the definition of $C_k(j,A)$ in Eq. (7), the model that has been adopted for AOTF transfer function is a simpler sum of a sinc-squared and a Gaussian, for a total of 7 parameters:

$$TF(v, v_0, I_0, w, I_G, \sigma_G, q, n) = \text{F}_{\text{sinc}} + \text{F}_{\text{gauss}} + \text{F}_{\text{cntnm}}$$

$$\begin{cases} \text{F}_{\text{sinc}}(v, v_0, I_0, w) = I_0 w^2 \dfrac{\left[\sin\dfrac{\pi(v-v_0)}{w}\right]^2}{\pi^2(v-v_0)^2} \\ \text{F}_{\text{gauss}}(v, v_0, I_G, \sigma_G) = I_G \exp\left[\dfrac{-(v-v_0)^2}{\sigma_G^2}\right] \\ \text{F}_{\text{cntnm}}(v, v_0, q, n) = q + n(v-v_0) \end{cases} \quad (8)$$

where:
- $v_0$ is the center of the AOTF transfer function (in cm$^{-1}$). The first estimate of this value is computed using the relation between $v_0$ and $A$ according to Eq. 2, using the prior coefficients;
- $I_0$ the sinc-squared amplitude;
- $w$ the location of the first zero-crossing of the sinc-squared function; the relation between $w$ and the sinc-squared FWHM is FWHM ≈ 0.886 $w$. However, in this case, FWHM is not straightforwardly the AOTF transfer function bandpass amplitude;
- $I_G$ the Gaussian amplitude;
- $\sigma_G$ the Gaussian standard deviation;
- $q$ and $n$ the continuum offset parameters

The set of $G_{h,h=0,1,2}$ values of Eq. (2), are computed by polynomial fit of $v_0$ vs. the corresponding AOTF frequency $A$ at which the $L_{d,k}(A)$ function peaks. Results are illustrated in Figure 8 for SO and Figure 9 and LNO, fitting all the data from the 30 SO and 28 LNO available miniscans. It is evident that the two sets of coefficients are not equal, since the two

AOTFs have not the same size and properties (Neefs et al., 2015). Residuals reported in the bottom panels show that the maximum calibration error is ~5 cm$^{-1}$ for SO and ~3 cm$^{-1}$ for LNO.

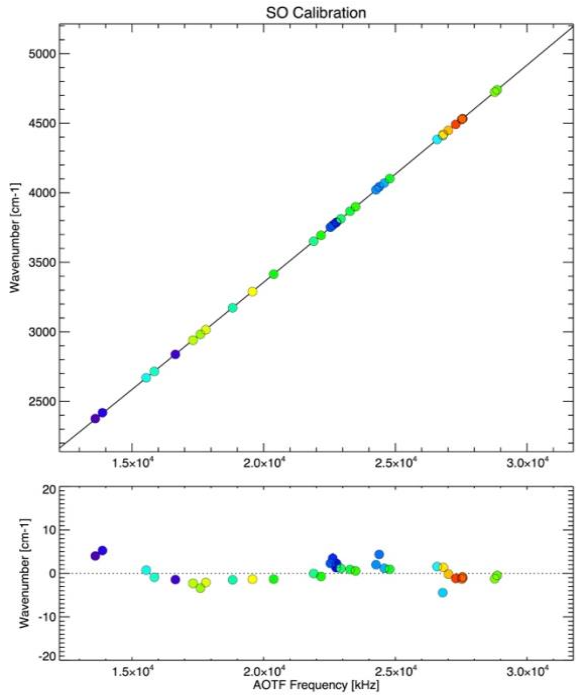
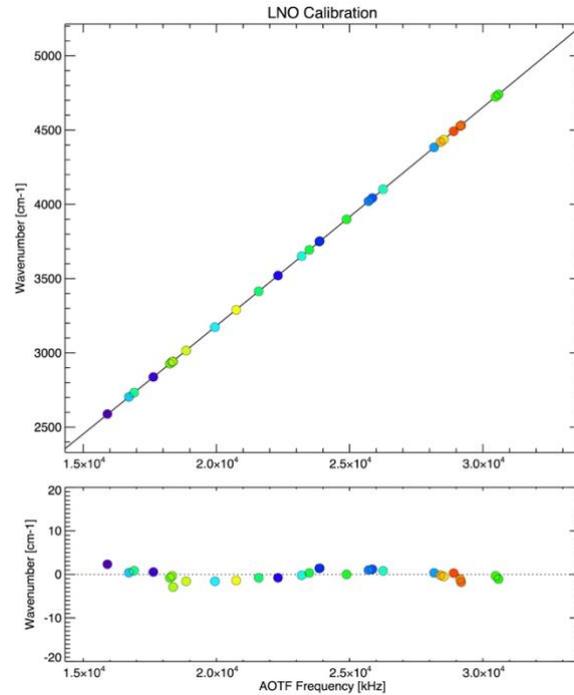

*Figure 8 - Results for all the available SO miniscans on the four dates.*

*Figure 9 - Results for all the available LNO miniscans on the four dates.*

| INSTRUMENT | DATE | $G_0$ | $G_1$ | $G_2$ |
|---|---|---|---|---|
| SO | 22$^{nd}$-27$^{th}$ Nov. 2016 | 313.91768 | 0.1494441 | 1.340818·10$^{-7}$ |
| LNO | 22$^{nd}$-27$^{th}$ Nov. 2016 | 300.67657 | 0.1422382 | 9.409476·10$^{-8}$ |

*Table 2 - List of the wavenumber vs. AOTF frequency calibration coefficients obtained using miniscans of selected dates.*

The color of the points indicates the miniscan in which the corresponding solar line has been observed and tracked, and so might be associated with various instrumental conditions (i.e., temperatures). Residuals from various miniscans seem not to exhibit any particular pattern in this sense, hence it is reasonable to state that the AOTF tuning relation does not depend on parameters such as instrumental temperature. The coefficients are reported in Table 2.

In order to characterize the AOTF transfer function, we analyzed the depth of a selected set of solar lines across the miniscans for a broad range of frequencies. This correspond to all the lines for which the convergence conditions are satisfied; for each of them, we subtract the continuum function derived from the fit, re-center the $L_{d,k}(v)$ around zero, and perform a fit to the AOTF line shape. Examples are illustrated in Figure 10 for two solar lines measured with the SO channel, at 3172 cm$^{-1}$ (left) and 3289 cm$^{-1}$ (right). The fitting parameters sets are provided in Table 3, where the two parameters correspond, respectively, to the position of the first zero-crossings of the sinc-squared component of AOTF function ($w$), and an enhancement factor scaling the sidelobes amplitude with respect to the normal sinc-squared function.

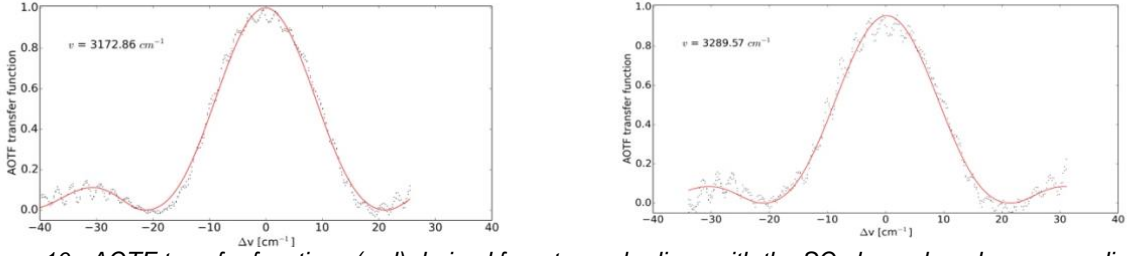

Figure 10 - AOTF transfer functions (red) derived from two solar lines with the SO channel, and corresponding fitting models (red).

| INSTRUMENT | SOLAR LINE FREQ. | $w$ | $I_{sidelobes}$ |
|---|---|---|---|
| SO | 3172.86 cm$^{-1}$ | 21.4381 | 1.5393 |
| SO | 3289.57 cm$^{-1}$ | 21.3935 | 1.3722 |

Table 3 - Retrieved coefficients for the shape of the AOTF transfer functions in Figure 10.

## 4 Spectral continuum modeling

The elements outlined so far are useful both to build a model for the continuum in the NOMAD spectra, and then to quantify the contribution to the total observed flux from the nearby orders, which is due to the side lobes of the AOTF transfer function.

The measured continuum shape is the result of the incoming continuum (Sun, Mars) modified by the combined shape of the AOTF transfer function, and the grating efficiency due to blazing. The removal of the instrumental effects is an essential part of the calibration process, especially in the interpretation of LNO nadir data, for which no reference measurements are possible. For SO occultation spectra, solar spectra taken before ingress begins (for egress, after it ends) provide the reference measurements.

Before introducing the detailed calculation of the spectral continuum, it is necessary to define the blaze function for the grating, for which we adopt a sinc-squared model (Engman and Lindblom, 1982):

$$F_{blaze}(p, p_0, w_p) = w_p^2 \frac{\left[\sin\frac{\pi(p-p_0)}{w_p}\right]^2}{\pi^2(p-p_0)^2} \quad (9)$$

where $p$ is the pixel number (from 0 to 319), $p_0$ is the center of the function in pixel units, and $w_p$ is the width of the blaze function. This last parameter is the equivalent in pixels of the Free Spectral Range, which is defined by Eq. 1. Both SO and LNO are characterized by a moving blaze function, whose central pixel, namely $p_0$, slightly moves with the diffraction order $m$, according to the following equation, derived by the analysis of fluxes at different orders:

$$p_0(m) = 150.80 + 0.22 \cdot m \qquad (10)$$

A contour plot of the blaze function as a function of diffraction orders is provided in Figure 11. The intensity of the blaze function, together with that of the AOTF transfer function, modulates in each pixel the intensity of the continuum. Thus, the total observed flux will be the sum of the different contributions coming from the central order (defined by the AOTF) and the nearby orders. If the spectrum is acquired at an AOTF frequency $A$ (the independent variable), whose corresponding wavenumber $v_0$ falls in the diffraction order $m$ the continuum takes the subsequent form, here denoted with $PEC(A)$ (standing for *Partial Elements Continuum*, a function of the AOTF frequency):

$$PEC(A) = \sum_{j=m-\Delta m}^{m+\Delta m} PE(j) = \sum_{j=m-\Delta m}^{m+\Delta m} AOTF(A, \boldsymbol{v_j}) \cdot F_{blaze}(j, \boldsymbol{v_j}) \cdot gain(j) \qquad (11)$$

where $AOTF(A, \boldsymbol{v_j})$ stands for the AOTF function at the AOTF frequency $A$, calculated on the spectral grid $\boldsymbol{v_j}$ of the diffraction order $j$, $F_{blaze}(j, \boldsymbol{v_j})$ is the blaze function of the diffraction order $j$, and $gain(j)$ is the spectral average throughput in the order $j$. However, if one considers a small amount $\Delta m$ of orders around the central order $m$, the gain is not expected to vary dramatically, hence it can be set equal to 1 for every order, and the resulting normalized values of the $PEC(A)$ function will already represent the continuum of the NOMAD spectrum,

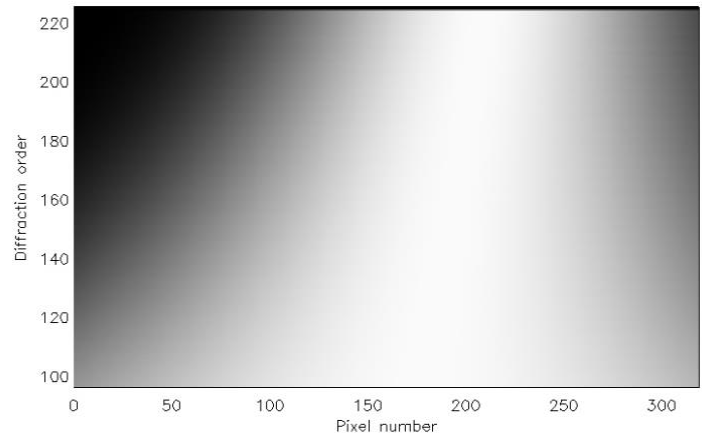

*Figure 11 - Map of the relative intensity of the blaze function according to the order number and the pixel, based on Eq. (10).*

normalized to the unity as well.

The derivation of the continuum according to Eq. (11) enables also to quantify the amount of flux coming from the central and the nearby orders as the displacement between the peak of AOTF transfer function and the central pixel of the central order varies. A summary of this calculation is presented in Table 4 (both for SO and LNO), in which it is shown that the contribution from nearby orders can reach ~50% of the total flux for SO and ~45% for LNO for higher orders (this is also due to the significant overlap between the spectral intervals of nearby orders, increasing with order number). Also, the sum of the contribution from central and nearby orders puts in evidence that taking into account ±3 orders around the central one is equivalent to consider between 97.6% and 98% of the total flux for SO, and more than 99% for LNO.

| SO ORDER | NEARBY ORDERS | AOTF CENTERED | AOTF FREQ. ± 20 kHz | AOTF FREQ. ± 50 kHz |
|---|---|---|---|---|
| 120 | **Central** | **0.7722** | **0.7359** | **0.5804** |
|  | 1st nearby | 0.1579 | 0.1896 | 0.3292 |
|  | 2nd nearby | 0.0346 | 0.0381 | 0.0492 |
|  | 3rd nearby | 0.0144 | 0.0155 | 0.0187 |
| 160 | **Central** | **0.6616** | **0.6386** | **0.5348** |
|  | 1st nearby | 0.2627 | 0.2816 | 0.3738 |
|  | 2nd nearby | 0.0412 | 0.0431 | 0.0485 |
|  | 3rd nearby | 0.0132 | 0.0148 | 0.0197 |
| 200 | **Central** | **0.5417** | **0.5376** | **0.5020** |
|  | 1st nearby | 0.3705 | 0.3753 | 0.4132 |
|  | 2nd nearby | 0.0487 | 0.0477 | 0.0442 |
|  | 3rd nearby | 0.0157 | 0.0162 | 0.0179 |

| LNO ORDER | NEARBY ORDERS | AOTF CENTERED | AOTF FREQ. ± 20 kHz | AOTF FREQ. ± 50 kHz |
|---|---|---|---|---|
| 120 | **Central** | **0.8171** | **0.7859** | **0.6443** |
|  | 1st nearby | 0.1557 | 0.1851 | 0.3204 |
|  | 2nd nearby | 0.0139 | 0.0152 | 0.0196 |
|  | 3rd nearby | 0.0056 | 0.0060 | 0.0072 |
| 160 | **Central** | **0.7019** | **0.6825** | **0.5899** |
|  | 1st nearby | 0.2667 | 0.2851 | 0.3741 |
|  | 2nd nearby | 0.0183 | 0.0185 | 0.0194 |
|  | 3rd nearby | 0.0060 | 0.0063 | 0.0073 |
| 200 | **Central** | **0.5858** | **0.5787** | **0.5386** |
|  | 1st nearby | 0.3781 | 0.3856 | 0.4258 |
|  | 2nd nearby | 0.0212 | 0.0207 | 0.0197 |
|  | 3rd nearby | 0.0070 | 0.0069 | 0.0069 |

*Table 4 - Summary of the contribution to the total observed flux from central and nearby orders according to the order (first column) and the displacement between the AOTF transfer function peak and the central pixel. Displacement is in AOTF frequency units (kHz). Top table is for SO, bottom for LNO.*

The formulation of continuum expressed through Eq. (11) can be straightforwardly used to compute the radiance as observed by NOMAD, using that as a weighting function for the radiance of central and nearby orders. Hence, observed radiance will be expressed as:

$$R(A, \bm{v_m}) = \sum_{j=m-\Delta m}^{m+\Delta m} AOTF(A, \bm{v_j}) \cdot F_{blaze}(j, \bm{v_j}) \cdot gain(j) \cdot R(j, \bm{v_j}) \qquad (12)$$

which can be considered the very final product of this calibration process, embodying all the elements needed to simulate, interpret and normalize NOMAD spectra in any observation geometry. An important aspect to note here is that this equation includes both the continuum introduced by the instrument and the target; in fact, all the effects related, e.g., to the planetary surface temperature, reflectance (mineralogy), and aerosol extinction and scattering are all embodied within the signal terms $R(j, v_j)$.

# 5   CH$_4$ sensitivity in Solar Occultation geometry

With the availability of a complete model for the instrument, it is now possible to study the effective sensitivity of NOMAD with respect to its main target species. As already briefly mentioned in Vandaele et al. (2015) and studied more in depth in Vandaele et al. (2018), one of the main limiting factors for trace gases detection is the presence of aerosols, namely dust and water ice, constantly present in the Martian atmosphere. In previous studies (Robert et al., 2016), this effect was taken into account by a conservative estimate of the Signal to Noise Ratio (SNR) of the instrument. Furthermore, NOMAD spectra are affected by different sources of noise (Thomas et al., 2016) which impact in different ways NOMAD observations, mainly according to target and, consequently, geometry.

In this work, we investigate how different scenarios of aerosols and configurations of the NOMAD infrared channels impact the sensitivity to trace species, specifically methane. The calculation of sensitivities is conceptually structured in three passages: first, the solar or Mars irradiance at NOMAD is computed by radiative transfer simulations, in which both the occultation height and the aerosol load are made to vary individually. Together with irradiances at NOMAD, an estimation of the noise (and consequently of the SNR) is provided. Then, the depth of CH$_4$ signatures in the spectra is compared to the noise to estimate a CH$_4$ effective SNR, from which sensitivities are computed.

Radiative transfer simulations were created using the Planetary Spectrum Generator (PSG, Villanueva et al., (2015), Villanueva et al., (2018)). PSG is an online tool for synthesizing planetary spectra (atmospheres and surfaces) for a broad range of wavelengths (0.1 µm to 100 mm, UV/Vis/near-IR/IR/far-IR/THz/sub-mm/Radio) from any observatory (e.g., JWST, ALMA, Keck, SOFIA), any orbiter (e.g., MRO, ExoMars TGO, Cassini, New Horizons), or any lander (e.g., MSL). This is achieved by combining several state-of-the-art radiative transfer models, spectroscopic databases and planetary databases (i.e., climatological and orbital).

The radiative transfer relies on the most updated spectral line compilations, gathered in the newly released HITRAN 2016 database (Gordon et al., 2017). For water and its isotopologues, a specific line compilation tailored on a CO$_2$-rich atmosphere is used (Villanueva et al., 2012). Radiative transfer can be done either via line-by-line calculations, or using a much faster k-correlated approach; in the case of NOMAD, given the high resolving power, the line-by-line approach has to be preferred, also because of the narrow spectral interval (~30 cm$^{-1}$ at most) covered by each diffraction order.

The computation of the scattering and overall extinction effects due to atmospheric aerosols is based on a Martian scattering model (Smith et al., 2013). In the present work, dust and ice are treated together, and their total optical depth is indicated at 3.3 µm, to refer directly to the aerosol opacity at the wavelengths at which we are conducting this analysis.

The PSG contains also a module for computing the noise for quantum and thermal detectors. The primary source of noise in solar occultation is the actual photon noise originated from the Sun signal, while in nadir measurements, the sensitivity is mainly limited by thermal background (TB is in [e-/s]: $1.5 \cdot 10^7$ at 263 K, $3 \cdot 10^7$ at 273 K, to $4 \cdot 10^7$ at 283 K). Due to a late removal of the active cooler in the NOMAD instrument during construction, the instrument is now insufficiently cooled, leading to particularly reduced sensitivities in nadir than originally planned. Other sources of noise include read-out-noise (1000 e-/s) and the dark current of the detector.

## 5.1 Input profiles and instrumental setting

The spectral interval spanned by NOMAD IR channels contains two main absorption bands of $CH_4$: the most prominent one is the $\nu_3$, associated with a stretching mode, whose Q-branch is centered at 3018 cm$^{-1}$ (3.3 µm). Also, a much weaker $CH_4$ absorption occurs at 2.4 µm ($\nu_1+\nu_4$), but is heavily masked by $CO_2$. Hence, this analysis is focused on the first one.

Based on the calibration results, the Q-branch of the $\nu_3$ $CH_4$ band falls within the diffraction order 134 (with the note that the Q-branch is offset to one side of order 134, not centered on it; Figure 12, 13a). Hence, radiative transfer simulations have been performed to compute radiances for orders from 131 to 137. For the sake of clarity, the single partial elements of Eq. (11) for orders 131-137 are represented in Figure 12, which provides a quick look at the relative contribution of nearby and central orders. In this case, orders 131-133 and 135-137 contribute 23% of the total flux observed by NOMAD. This is not a simple side note, because also the spectral intervals of orders 133 and 136 in particular contain strong $CH_4$ lines, and some orders contain weak lines of $^{18}O^{12}C^{16}O$ ($\nu_2+\nu_3$) and all orders contain strong lines of Mars $H_2O$ (mainly the $2\nu_2$ band). Solar lines also are present in this region.

As already stated, in Solar Occultation observations the Sun disk illuminates, in average, only the central 23 rows of the detector. This number changes slightly with the distance from the Sun. Each of the rows contains information about a different occultation path (slant height, and line-of-sight extent). In order to evaluate the information content of NOMAD spectra at the maximum spatial resolution, simulations are performed for the radiance as observed by the single detector row, without any spatial binning. Every detector pixel samples a 1 arcmin angle along the slit, and when considering that the slit-width for SO is 2 arcmin, this leads to an observational field-of-view (FOV) of 2 arcmin$^2$. Besides radiances, the corresponding noise has also been synthesized.

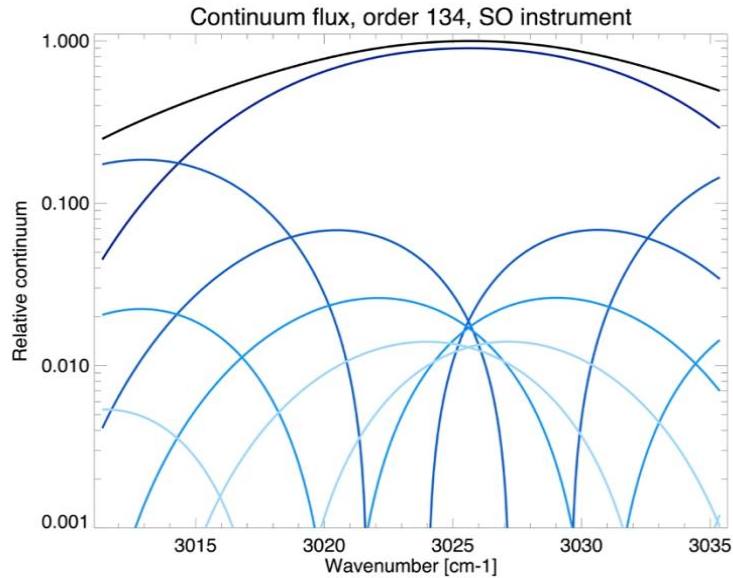

*Figure 12 – Relative flux contribution from the central order (darkest line) and nearby orders (lightest lines are for farthest orders). Black line is the total normalized continuum.*

Radiative transfer calculations have been performed assuming a single model for atmospheric temperature and gaseous concentration profiles, computed using the LMD General Circulation Model, v5.3 (Millour et al., 2015). The chosen profile is representative of a mild, relatively warm season ($L_S$ = 180º) in the Equatorial zone. Though this could initially sound like a limitation, for the purpose of this analysis the two independent variables are aerosol load and observation height, which are both varied, keeping fixed only the profiles for atmospheric temperature and gaseous abundances.

The atmospheric model is structured in 30 layers with $p_0$ = 5.00 mbar the pressure at ground, and the Top of Atmosphere (TOA) at 1.2 x 10$^{-7}$ mbar (90 km altitude). Radiative transfer calculations have been done from the surface up to 75 km of height, at steps of 3 km. For the sake of sensitivity calculation, $CH_4$ is assumed to be vertically well-mixed in the atmosphere, with a constant volume mixing ratio of 1 ppbv. Even if the choice of a uniform vertical profile might seem not suitable for the sake of describing the complex dynamics of the Martian atmosphere, a uniform vertical profile makes the results of the sensitivity calculations independent of the particular phenomenon (e.g. surface release, high atmosphere chemistry) that generates and depletes methane, and provides an easy way to rescale the results according to the specificity of the actual vertical profile. In addition, the results obtained with this choice have the potential to characterize, together with sensitivity, also the minimum variations in $CH_4$ concentration that NOMAD can actually observe at each altitude.

Heights from 75 to 90 km have not been considered in the simulations, since atmospheric absorption becomes negligible at those occultation altitudes.

Six different scenarios for aerosol opacities are considered, corresponding to six values of nadir aerosol opacity at 3.3 µm: 0.015 (nearly-clear sky), 0.15 (low aerosol), 0.225 and 0.375 (moderate aerosol), 0.6 (high aerosol opacity) and 1.5 (aerosol opacity typical of huge dust storms). The vertical distribution of the aerosol is assumed to be well-mixed. The two opacities corresponding to moderate aerosol conditions are also representative of the global annual average of the sum of dust and ice opacities at 3.3 µm, which is obtained after rescaling the values at TES reference wavelengths (Smith, 2004, 2008). The six sets of radiative transfer simulations resulting from the six aerosol opacities will yield six different vertical profiles for $CH_4$ sensitivities.

## 5.2 Results and sensitivities

The PSG software has been used to synthesize spectra at a resolution needed to resolve the Martian line shapes (0.001 cm$^{-1}$), and later binned to the instrument spectral resolution. Then, within the pipeline to read and process the spectra, the spectra are convolved with the instrument convolution kernel ($\lambda/\delta\lambda$ ~ 19000), obtained from the calibration process. PSG is used to provide directly collected radiances for each order in quanta (photons), which are then converted to counts (ADU) through the detector gain (equal to 900 photons/ADU), and then added all together according to Eq. (12), to obtain directly the simulated NOMAD spectrum. In this exercise, the AOTF frequency is set equal to 17859 kHz, and is derived from the calibration process as the one that maximizes the flux of order 134 over the nearby orders. To mimic one of the possible setups of the instrument during a Solar Occultation measurement, simulations have been done considering 16 co-added exposures (accumulations) each one with an integration time of 0.01 s, yielding a total integration time of 0.16 s.

Results of the simulations are summarized in Figure 13. Each panel refers to a different aerosol load, and each spectrum corresponds to a different occultation height. The effect of aerosol extinction is very evident already at the minimum load level, in which aerosol cuts ~50% of the flux reaching the instrument from lower occultation slant heights with respect to that at TOA. As the aerosol load increases, the atmosphere becomes opaque starting from increasing heights. Typical, average global conditions across all the planet for all $L_S$ values are represented in panels c) and d), where nadir aerosol optical depth ranges between 0.225 and 0.375. In this case, the atmosphere looks to be totally opaque (transmittance equal to 0) for occultation heights below 10 km. The final simulation (panel f) represents a situation that typically occurs during regional dust storms of moderate intensity.

The most salient spectral features visible in these first results are those due to water vapor ($H_2^{16}O$) and solar lines, while the optical depth of $CO_2$ is negligible, since radiative transfer calculations have evidenced that it is much lower than the one corresponding to methane.

The evaluation of $CH_4$ sensitivities is based on the noise model embedded in PSG. Noise has been computed for every spectrum. It is expected that, as aerosol optical depth increases, noise will increasingly prevail over signal, yielding to $SNR \rightarrow 0$ for very high aerosol loads at low occultation heights. Sensitivities have been quantified by computing the ratio between the average noise (expressed in counts) in the spectral region 3015 – 3020 cm$^{-1}$ - where $CH_4$ absorption is maximum - and the maximum depth of $CH_4$ features. This last quantity is computed as follows: if $T_{min}$ is the minimum transmittance of $CH_4$, occurring in the spectral channel with wavenumber $v_{min}$, and $C(v_{min})$ the continuum intensity (in counts) at $v_{min}$, the depth of $CH_4$ lines will be $(1 - T_{min})C(v_{min})$, and will be expressed in counts.

It is important to state again that the sensitivity is computed based on the intensity of the brightest $CH_4$ line in the order, meaning that not all the information content of the spectrum is exploited. As evidenced in the subsequent plots, indeed, even if the sensitivity is computed only on the brightest $CH_4$ line, the spectral interval 3010 – 3036 cm$^{-1}$ contains other bright methane lines (R0), together with the Q-branch. As a consequence, the sensitivities presented in this work are actually a conservative estimate of the real sensitivity that can be achieved, in this spectral domain, through spatial binning or using the whole spectral range, performing a full retrieval via damped Least Squares or Optimal Estimation approaches (Liuzzi et al., 2016; Villanueva et al., 2015).

Furthermore, we assume that systematics will restrict the continuum SNR to a maximum of 1000, hence we have capped the modeled SNR to 1000. Both of these conditions, make our estimates conservative, yet probably realistic considering our experience in analyzing high-resolution spectra of Mars. With the same outline adopted for the results depicted in Figure 13, the SNR$_{CH4}$ values are presented in Figure 14. At a first glance, these seem to indicate that SNR$_{CH4}$ ranges between 0 (low occultation heights, very high aerosol opacity), and 10 (very low opacity) for 1 ppbv of $CH_4$. This further perspective on results of simulations demonstrates that the usual aerosol loads that can be found in the Martian atmosphere will dramatically reduce the flux reaching the instrument when the occultation height is within the lower troposphere, significantly affecting the sensitivity to $CH_4$.

However, if rapid vertical transport occurs under some conditions (e.g., Viscardy et al., 2016, Holmes et al., 2017), the $CH_4$ mixing ratio could be significantly enhanced in the middle atmosphere. A constant vertical load then could imply a middle atmosphere mixing ratio much greater than the mean 1 ppbv mixing ratio assumed here.

A comprehensive chart showing methane sensitivities, computed as described above, is shown in Figure 15.

Each line of the plot corresponds to a different aerosol load (see panels in Figure 13 and Figure 14). While sensitivities are highly differentiated in the lower part of the atmosphere, they tend to converge to values independent of aerosol abundance, for occultation heights higher than 30 km. On the one hand, this happens because of the upper limit of 1000 imposed to SNR, which makes the $CH_4$ sensitivity equal for any aerosol load in the upper atmosphere; on the other hand, as already explained, an increasing amount of dust suppresses steeply the signal in the lower atmosphere, making the $CH_4$ sensitivities diverge as occultation height decreases. Overall, the present analysis strongly indicates that sensitivity values are within the interval 0.10 – 1.5 ppbv in the altitude range 0 – 30 km for the lower aerosol load, reaching tens of ppbv for higher aerosol concentrations in the lower atmosphere. For average conditions, according to Mars climatology, ($\tau$ ~0.3, average between the two green lines), a sensitivity of ~0.65 ppbv at ~15 km of altitude can be inferred.

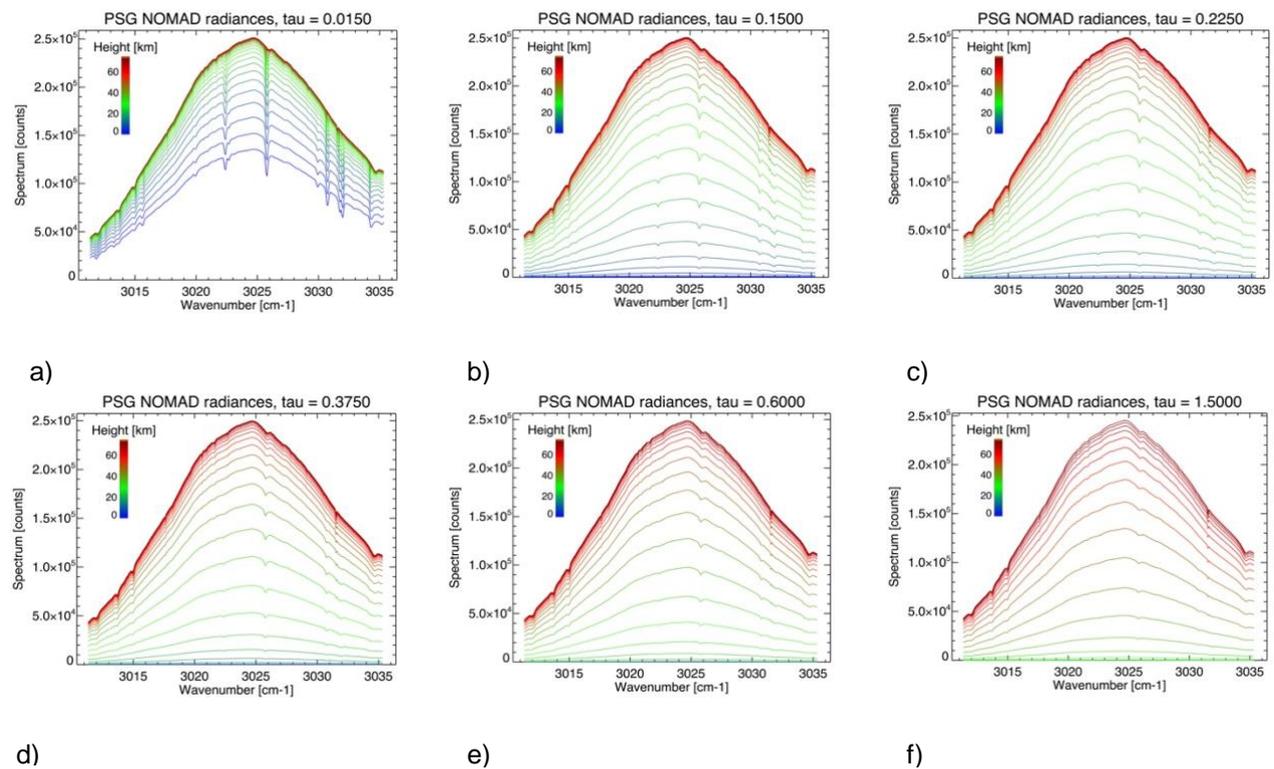

*Figure 13 – Ensemble of radiative transfer simulations results for different aerosol loads: panel a) $\tau = 0.015$, b) $\tau = 0.15$, c) $\tau = 0.225$, d) $\tau = 0.375$, e) $\tau = 0.6$, f) $\tau = 1.5$. The height, shown as insert (color bar), is the slant height of the line-of-sight viewed by a single pixel (row) of the NOMAD detector looking through the Mars atmosphere in SO mode. A uniform vertical mixing profile (1 ppbv) is assumed for $CH_4$ in each simulation.*

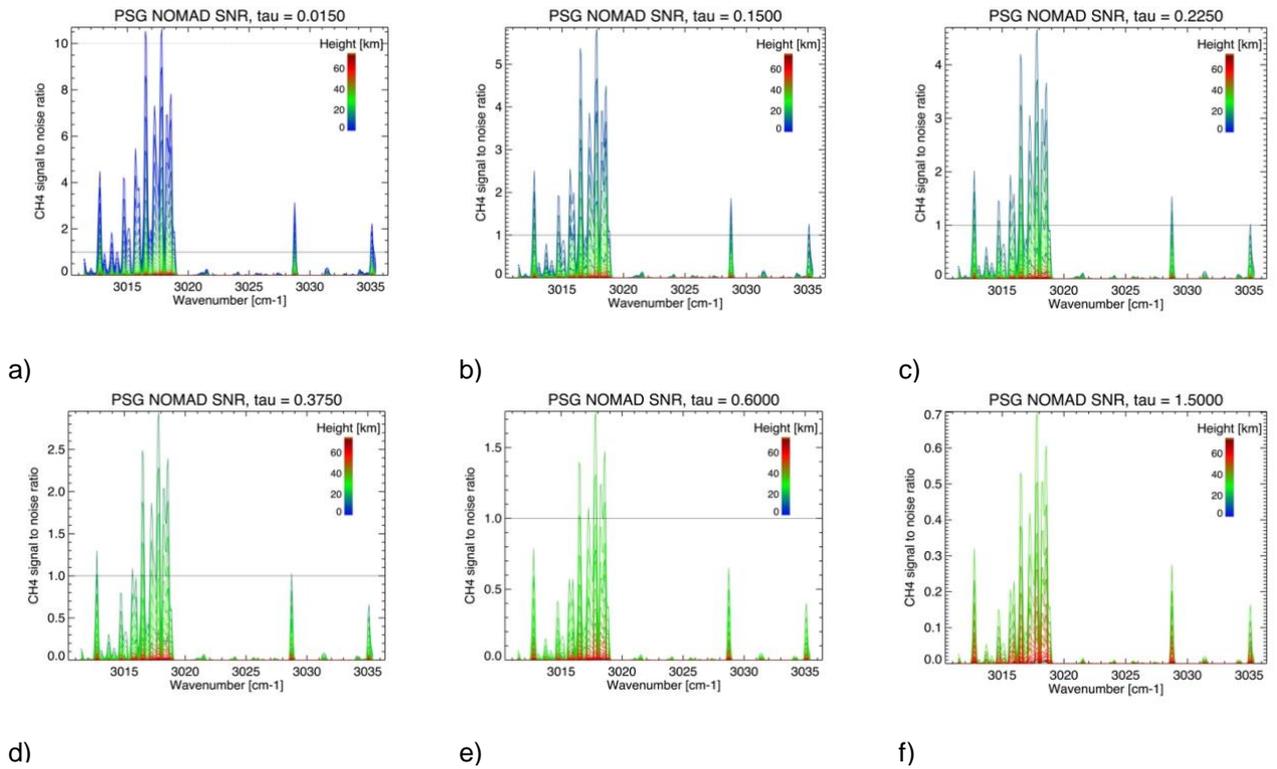

a)   b)   c)
d)   e)   f)

*Figure 14 – Values of CH$_4$ Signal to Noise Ratio computed for different aerosol opacities. Values of opacities and color scales are the same as in Figure 13.*

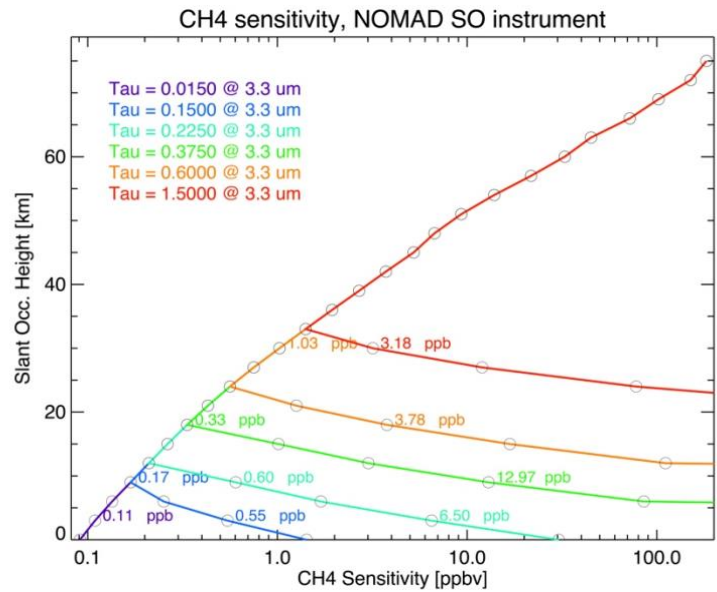

*Figure 15 – CH$_4$ sensitivity computed for different aerosol loads and occultation heights. Each solid line corresponds to a distinct aerosol opacity in the nadir direction (see inset caption). The slant altitude along the line-of-sight for each NOMAD pixel (spectrum) is shown on the y-axis. For more immediate readability, sensitivities are indicated at heights equal to 3, 9, 18 and 30 km.*

## 5.3 Discussion

From a quantitative point of view, the sensitivities resulting from this analysis reveal a relatively different scenario from that outlined in previous studies (Robert et al., 2016; Vandaele et al., 2015). Those works claimed a $CH_4$ detection limit of 0.025 ppbv, which is ~4 times better than the best values obtained in this work (0.10 ppbv at the lowest occultation height and for the lowest aerosol load, see Figure 15).

This difference is based on several factors: first of all, the results presented in (Robert et al., 2016; Vandaele et al., 2015) considered a single occultation slant height value (20 km); second, the SNR value here is locked to a maximum of 1000, compared to 2000 or more in previous studies; this negatively impacts (by a factor of at least 2) our work's best estimate of sensitivity. Third, both the actual signal intensity – due to $PEC(A)$ function – and the mixing of central and nearby orders, are introduced only in the current analysis. Order mixing, in particular, introduces "foreign" flux from nearby orders at a level of 30% in the spectral interval 3015 – 3020 $cm^{-1}$, where methane absorption is maximum. This reduces (by the same amount) the actual sensitivity of NOMAD to $CH_4$ in this spectral region.

Finally, previous simulations did not fully consider the impact of aerosol extinction, contrasting with the present work. The recent results presented in Vandaele et al., 2018 deal with this aspect in the calculation of detection limits for $CH_4$ and other gases, and are in general agreement with those presented in this work. Vandaele et al. shows that dust opacity impacts the capabilities of NOMAD for detecting $CH_4$ at tangent heights below 20 km for moderate aerosol loadings (green line in Figure 15), and below 30 km for the highest loads (orange to red line). The retrieval sensitivities presented here are nevertheless generally lower, and it is due to the factors already mentioned above, other than aerosol extinction impact.

The picture of NOMAD capabilities is completed confronting the sensitivities derived in this work with $CH_4$ abundances observed in previous detections. In particular, the most recent observations by MSL Curiosity, located in Gale crater, reported a value for $CH_4$ background concentration of 0.2 to 0.7 ppbv (Webster et al. (2015), Webster et al. (2018)), which should be detected by NOMAD with low aerosol loads ($\tau < 0.3$), at altitudes between 6 and 20 km. On the other hand, the maximum amount of $CH_4$ reported in literature is that observed during the strong release of northern summer in 2003 (Mumma et al., 2009). In this case, Mumma and colleagues reported a peak abundance of 33 ppbv from ground-based observations. In detail, two peak abundances were reported at $L_S$ ~121 and $L_S$ ~155. In both cases, the column peak abundance was ~40 ppbv, and the major plumes were located around Nili Fossae, Terra Sabae and Syrtis Major. The first plume observed at $L_S$ ~121 had an extension of 1270 x 820 $km^2$, while the second one of 950 x 580 $km^2$. All these values are far greater than the large majority of the sensitivity values reported in Figure 15. Overall, the event reported in Mumma et al. corresponds to a global average of ~6 ppbv once $CH_4$ has gone through global diffusion, which is well within the NOMAD sensitivity at any altitude below 45 km for $\tau < 0.2$, above 10 km for $0.2 < \tau < 0.6$, and only above 20 km with higher aerosol opacities.

In summary, looking at previous methane detections, it is evident that NOMAD will be able to provide enhanced spectral, spatial and temporal information, even considering all the limiting factors – both instrumental and atmospheric – which will affect observations.

The conclusions we have inferred so far are reinforced by further, related observables. Using TES climatology for dust and ice (Smith, 2004), and their known extinction properties (Wolff et al., 2009), both dust and ice opacities at TES reference wavenumbers (respectively 1075 and 825 $cm^{-1}$) are linearly rescaled to that at 3018 $cm^{-1}$. For each aerosol opacity, we assigned the corresponding best value for a $CH_4$ sensitivity, which is derived by the results achieved so far, at the slant occultation height at which it occurs (see Figure 15).

We then created a seasonal map for the $CH_4$ sensitivity, shown in Figure 16b, together with the seasonal map of the total optical depth at 3018 $cm^{-1}$ (Figure 16a); their comparison conveys an immediate outlook on the capabilities of the NOMAD under different conditions.

It is important to mention that the TES optical depths were measured at 2 p.m. local time, while the occultations occur at the terminator. Models suggest that the ice cloud optical depth can undergo significant variations, with possibly more clouds forming at the time of the occultation, which will be typically colder than the afternoon local times sampled by TES (e.g. Daerden et al., 2010). Hence, the presented results will be likely impacted on by the true ice cloud optical depth at the terminator.

The poor sensitivity at $L_S \sim 235º$ is due to the occurrence of large regional dust storms in that season. The highest sensitivity to $CH_4$ occurs around $L_S \sim 120º$ and latitudes between 40º and 65º N; however, the expected time for the start of the science phase will be around $L_S \sim 150º$.

It is fair to point out that this is exactly the same location of the first detection of $CH_4$ reported in Mumma et al. (2009), making this region of particular interest for methane detection by NOMAD SO measurements. The other peak values of $CH_4$ sensitivity occur at $L_S \sim 50º$ at latitudes $\sim 20º$ S and $L_S \sim 310º$ at latitudes between 50º and 70º S. In all these cases, the highest sensitivity is ~0.2 ppbv at ~10 km of altitude. Furthermore, model studies showed that methane released from the surface will spread over the planet (and with altitude) in a time scale of the order of a few Sols, such that any emission event can be detectable from many locations on the planet and not necessarily in the source region (Viscardy et al., 2016). This will increase the capability and likelihood of NOMAD to detect methane after surface release events.

Overall, our study results imply a sensitivity to $CH_4$ around 0.6 ppbv or better, apart from dust storms (seasonal) or thick ice clouds (typically at the poles and Equator at $L_S \sim 100º$). All these results yield the same conclusion as earlier work: NOMAD is a key instrument for progressing our knowledge of $CH_4$ abundance and its characterization under most atmospheric conditions. In addition, an *a priori* temporal and spatial mapping of the most likely sensitivities is potentially very beneficial for optimizing observational planning. Our results further demonstrate the ability to quantify the vertical mixing ratios for $CH_4$, and can be easily extended to $H_2O$ and other trace gases. This capability is a first for any Mars-orbiting spacecraft, and is a central to any attempt to understand the roles of transport and active chemistry in controlling the creation and destruction of such species.

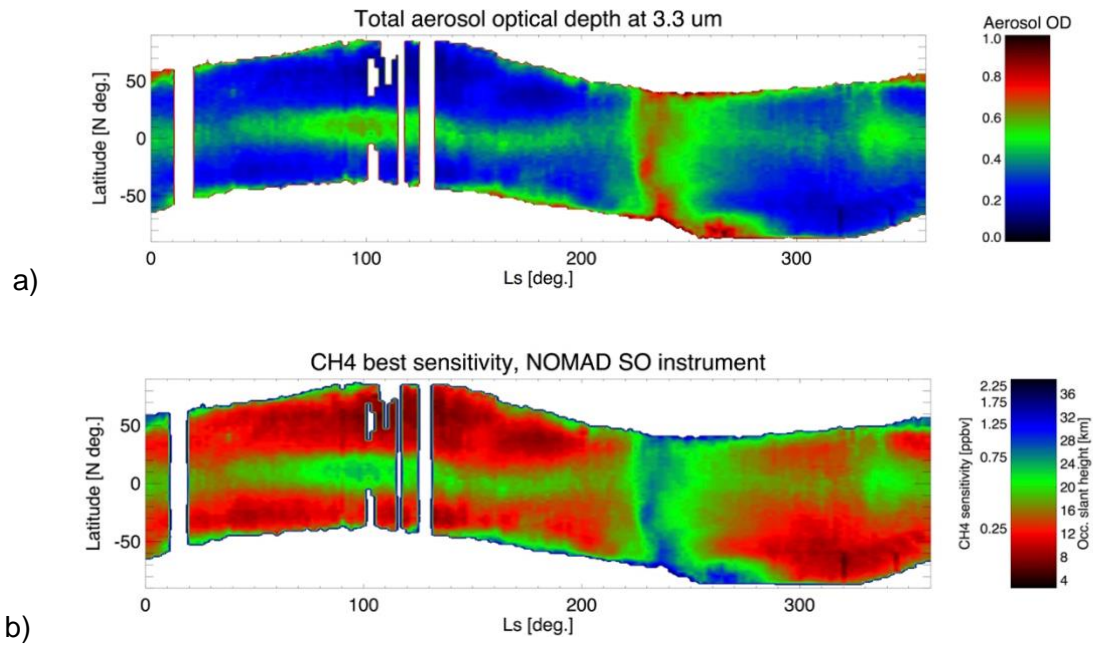

*Figure 16 – Seasonal sensitivities for $CH_4$ derived from the current analysis. a) Map of aerosol (dust + ice) total optical depth (OD) at 3.3 µm derived from TES products for MY24; b): best sensitivities occurring for every latitude and $L_S$, with the corresponding occultation slant heights. Below this height, the atmosphere becomes opaque because of aerosols and $CH_4$ sensitivity diverges. Blank spaces are missing data (in TES database) or sensitivities out of color scale range. Red color in this plot indicates higher sensitivity to $CH_4$.*

The values discussed so far can be further improved by binning the data spatially and spectrally. The sensitivities calculated before were derived considering a value for the SNR of a single detector row equal or lower than 1000 for a single altitude, yet by binning more rows (lower spatial resolutions) one can achieve higher SNRs. A further improvement factor can be obtained by spectral binning the data, specifically in the spectral interval 3013 – 3020 cm$^{-1}$, where $CH_4$ Q-branch signal would be dominant. This allows to exploit the entire information content of the data as far as $CH_4$ is concerned, whilst the sensitivities calculated before were based on the brightest line in the order. By contemplating the whole Q-branch of methane, we expect a factor ~8 improvement by spectral binning the data. Therefore, the detection sensitivity at 10 km for the whole Q-branch on a clear atmosphere for a spatial binning of 4 rows would correspond to ~ 0.5 ppbv / (8*√4) = 30 pptv (3-sigma), in relative agreement with the estimates presented in Vandaele et al., 2018.

# 6 CH$_4$ sensitivities in Nadir geometry

## 6.1 Methods and results

The method outlined for Solar Occultation geometry is easily extended to estimate the sensitivity of LNO channel in Nadir-looking geometry. Although the general guidelines for the analysis are essentially unchanged, the different targets and geometries imply some variations to the method, which are detailed here:

- LNO channel in nadir geometry is equipped with a much larger entrance slit (4 x 150 arcmin) and greater aperture diameter (27 mm against 20 mm for SO), to maximize SNR. As a result, during a nadir observation, 144 rows of the detector are illuminated (against 23 in SO); thus, the input beam FWHM is re-scaled to 27.08 arcmin for LNO nadir mode (vs. 0.8 arcmin for SO mode).
- Every nadir acquisition is characterized by an integration time much longer than for SO. We adopt a total integration of 12 s, corresponding to 20 accumulated measurements, each lasting 0.6 s.
- Since the signal detected during a nadir observation is orders of magnitude smaller than that during a Solar Occultation, simulations have been performed using a CH$_4$ concentration of 10 ppbv, instead of 1 ppbv.
- Unlike a Solar Occultation, the parameters that actually affect signal intensity in nadir observations in the near infrared are solar illumination, Solar Zenith Angle (*SZA*) and surface albedo. To verify this last point, some side simulations have been performed with the minimum and maximum aerosol cases. If the other properties of surface and atmosphere are left unaltered in the two cases, the difference in radiance is less than 50%. On the contrary, both *SZA* and surface albedo can impact the signal up to its maximum value. For this reason, simulations have been carried on with a single aerosol optical depth value ($\tau = 0.375$), which is a fair representation of global spatial and temporal average dust load.
- LNO instrument is typically operated at a lower temperature than SO. Thus, we have assumed an operating temperature T = 257 K for wavenumber and continuum calculation, using our calibration scheme. We set the AOTF frequency to 18927 kHz, the optimal value according to LNO calibration for order 134.
- Noise is calculated directly by PSG and its embedded noise model, but in this case, no upper limit is imposed to the computed SNR, which typically will have values on the order of ~100 (consistent with previous findings and studies, i.e. Robert et al., 2016).
- Differently from solar occultations, nadir observations are heavily affected by aerosol scattering, since in nadir geometry the instrument is pointed towards a diffuse source of radiation. In our calculations, we take into account these effects by full scattering calculations embedded in the PSG. By trial and error, we have found that 4 stream pairs and 16 Legendre polynomials to compute the phase function grant a precision better than 1% in the computed flux at 3.3 μm.

When operated in nadir geometry, the LNO channel is tailored to perform an extensive mapping of the planet surface every ~30 sols. What is more, since *SZA* and albedo are the main triggers of the flux reaching the instrument, the nadir sensitivity study is useful to complete the picture of NOMAD capabilities in the spatial domain, having already analyzed the capabilities of the instrument to depict the $CH_4$ seasonal (i.e. temporal) cycle with SO observations.

For simulations, we used the same atmospheric average input profile used for SO. To explore the sensitivity of the instrument at different *SZA* values, simulations have been conducted with *SZA* between 0 and 90 degrees, at steps of 5 degrees. In radiative transfer simulations, albedo is assumed to be constant and equal to 0.254, which corresponds to the case of a bright surface, whose definition is consistent to that given in Thomas et al. (2016).

Results of simulations, and corresponding $SNR_{CH4}$ values are reported in Figure 17. In this case, the $SNR_{CH4}$ spectral peak varies between 0 and 0.22. This clearly indicates that, to achieve a continuous mapping of methane concentration on Mars, spatial averaging of spectra will be needed. Using the same scaling of $SNR_{CH4}$ used for SO, it is possible to quantify the NOMAD sensitivity to $CH_4$ for each *SZA*. That relationship is shown in Figure 17c.

The graphics of Figure 17 clearly show that NOMAD sensitivity to $CH_4$ rapidly degrades for *SZA* greater than 50º, and it is between 35 and 50 ppbv for *SZA* < 50º. In practice, however, high *SZA* observations are limited during science operations, as LNO channel is switched on/off.

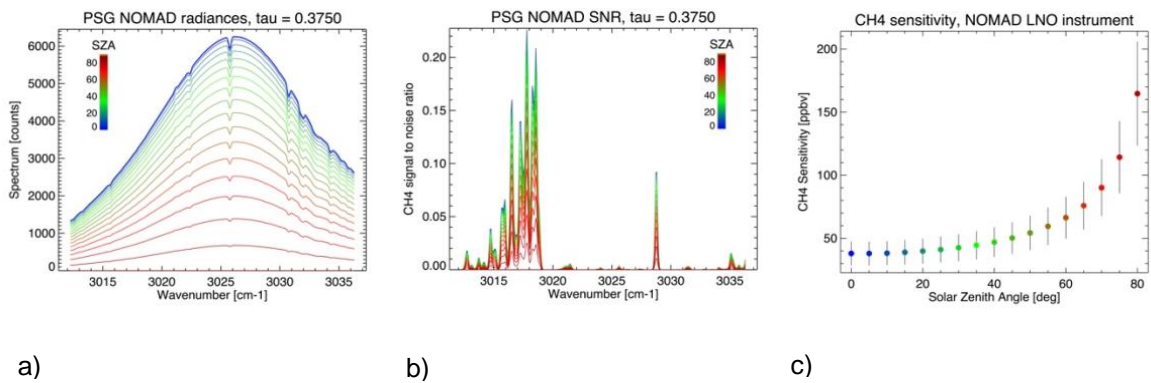

a)            b)            c)

*Figure 17 – a) LNO nadir simulations in the spectral interval covered by order 134. The spectral shape is calculated using Eq. 13. Color scale indicates the SZA; b) Values of $CH_4$ Signal to Noise Ratio corresponding to the spectra in panel a); c) Corresponding NOMAD sensitivities to $CH_4$ for each SZA value Error bars are referred to a ±25% variability of sensitivity, corresponding to the signal variation due to the variability of aerosol load (see above). For better readability, color scale is the same as left panels.*

## 6.2 Discussion and geographical mapping

The sensitivity values differ from those presented in Vandaele et al., 2018, where $CH_4$ detection limit ranges between 11 and 13 ppbv. This difference is mainly due to the SNR considered: calculations in the work by Vandaele et al. were performed with a SNR ~100, compared to a value of ~50 of this simulation, arising from the continuum in the region around 3015 – 3020 $cm^{-1}$. The possibility of spectrally bin the data is also not contemplated in this analysis (see Section 5.4).

The conclusions drawn so far for LNO nadir observations make evident that it is not practical and not sustainable to use the single spectrum for $CH_4$ geographical and seasonal mapping. For this reason, the capabilities of LNO channel in nadir mode should be rather viewed in terms of averaging different acquisitions.

For the sake of this analysis, let us consider a number of acquisitions close to 100, which would allow to substantially improve the sensitivity to $CH_4$ by a factor $\sqrt{n}$ ~ 10 for the stacked spectrum, yielding a sensitivity $S_{0,CH4}$ = 3.5 ppbv, where $SZA$ = 0º, albedo $\rho_0$ ~ 0.25 and surface pressure $p_{s,0}$ = 5 mbar, which are the parameters used for our calculations so far.

As a reference, let us compute NOMAD sensitivity on a spatial grid with resolution $\Delta s$ = 5º x 5º, which correspond to $\Delta x$ ~ 300 km. To provide a realistic estimate of the temporal interval needed to accumulate this number of observations in each bin, it is proper to refer to the orbital parameters of TGO and spatial resolution of NOMAD (Neefs et al., 2015). In its final orbit for science operations, the spacecraft will cruise at $v$ = 3 km/s, hence every spatial box will, in average, be spanned in latitude in a time $\Delta t = \Delta x/v$ = 100 s. If each acquisition has an integration time of 12 s, each geographical bin will be spanned, in latitude, at most by 8 acquisitions per orbit. The orbit of the spacecraft is structured in such a way that the instrument will map the entire planet every 30 sols, with a spacing between two consecutive footprints of ~1º in longitude. In turns, this means that each 5º x 5º geographical bin will be mapped by 8 x 5 = 40 acquisitions every 30 days, hence $n$ ~ 100 acquisitions on each bin will be obtained, in average, every ~70 Sols. This is a realistic temporal sampling at which global $CH_4$ concentration maps can be derived by NOMAD LNO nadir data, and enables to trace possible seasonal trends.

To map the global sensitivity to $CH_4$ in nadir geometry, the value of $S_{0,CH4}$ can be linearly rescaled with $SZA$, $\rho$ and surface pressure $p_s$:

$$S_{CH4}(SZA, \rho) = S_{0,CH4} \cdot \frac{\rho_0}{\rho} \cdot \frac{1}{cos(SZA)} \cdot \frac{p_{s,0}}{p_s} \quad (13)$$

This equation assumes that the dust load is constant, and neglects the dependence of scattering effects on $SZA$. Moreover, it neglects the contribution of thermal radiation coming from surface and atmosphere. However, it has been verified that this contribution is negligible in this spectral interval during daytime observations, being around 5% of the total flux observed by NOMAD at high *SZA* values, and less than 1% for *SZA* < 50º.

An example of a geographical map of sensitivity obtained via Eq. (13) is reported in Figure 18. To show the performances of NOMAD in any illumination condition, it has been chosen to compute sensitivity at $L_S$ = 90º (northern summer solstice), with *SZA* = 0º at East longitude = 0º. The albedo map used for calculations is that shown in the top panel of Figure 18, and is derived from the Multispectral Reduced Data Records of CRISM (Murchie, 2006). The map nicely shows the combined effect of *ρ* and *SZA* over the instrument sensitivity. It is interesting to note that, even though *SZA* values are larger at high latitudes, bright surfaces are able to keep the sensitivity around 7 ppbv even in part of these regions.

Overall, results indicate that in ~13% of geographical bins (321 over 2592) NOMAD has a column sensitivity better than 7 ppbv, and in 150 of them better than 5 ppbv, mostly concentrated between 10º and 50º N in latitude. The albedo map indicates that better sensitivity is expected in the northern hemisphere, where most bright regions are found.

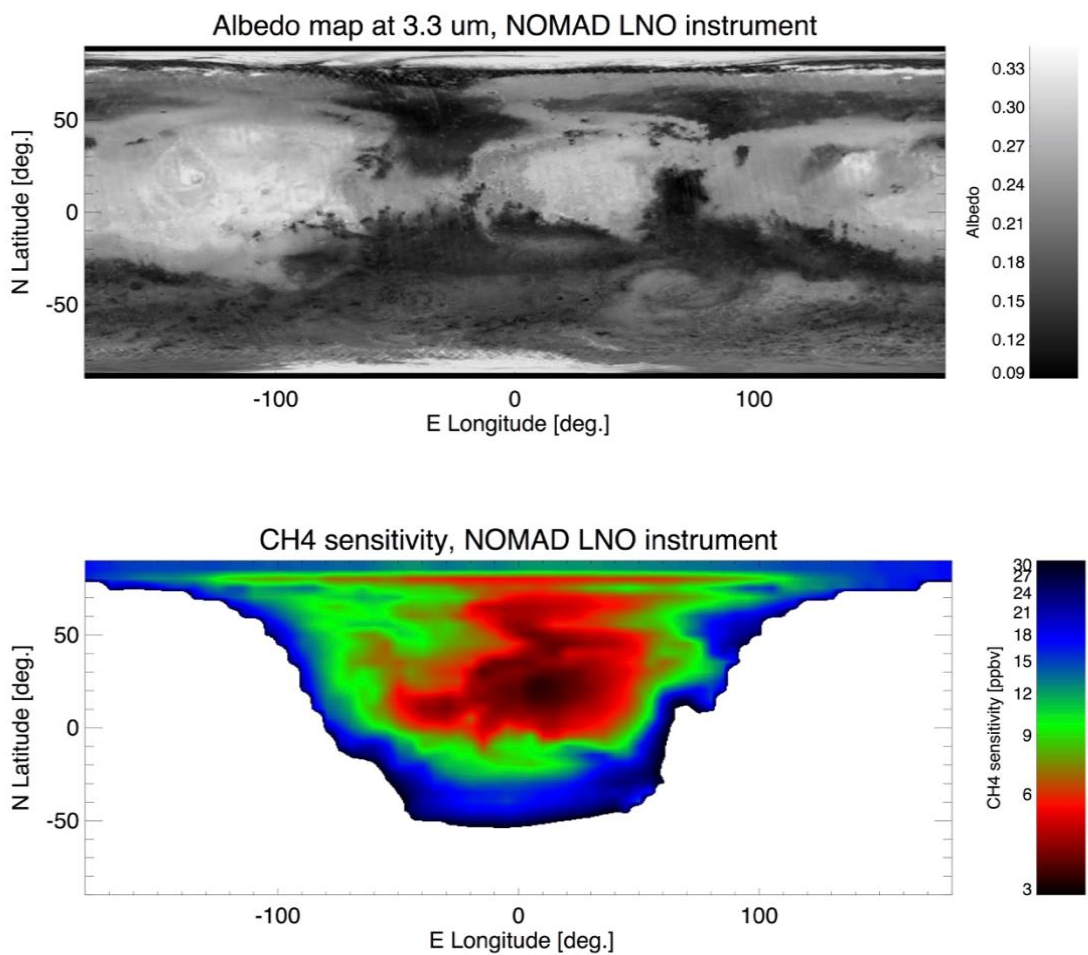

*Figure 18 – Top panel: albedo map used for sensitivity calculation (this is not the geographically binned version). Bottom panel: computed NOMAD CH₄ sensitivity in nadir geometry at $L_S$ = 90º. Blank pixels are characterized by SZA > 80º or by sensitivity values worse than 30 ppbv.*

# 7  Conclusions and remarks

This work has been structured to cover two essential aspects in characterizing the perspectives offered by NOMAD in bringing new science about trace gases in the Martian atmosphere. On the one hand, it has been intended to pursue a complete and accurate calibration of the instrument single components, which are essential to the second step, in which these elements are used to get information about the sensitivity of the instrument to a particular physical parameter in different atmospheric conditions.

The main product of the calibration process explained in this work is the elaboration of a model for the spectra as seen by the instrument. This model can reproduce the measured spectral continuum, and it reveals the way in which different diffraction orders mix to produce the measured spectrum. It has been shown that a fraction of the total flux (between 15% and 50%) in a targeted order is actually contributed by nearby orders, and that this fraction increases with order number (and AOTF frequency).

The model has been applied to the problem of detection of $CH_4$ with NOMAD both in Solar Occultation and Nadir geometries. The analysis carried on in this work has clearly shown that aerosol in the Martian atmosphere dramatically impact the detectability of $CH_4$ (and likely of many other trace species) in SO geometry, especially at occultation altitudes lower than ~10 km. Notwithstanding this limitation, NOMAD capabilities in this context look to be often better than the remote measurements of $CH_4$ reported in the literature, hence the instrument is robust in this sense.

As far as LNO nadir observations are concerned, it has been shown how the instrument will be able to produce a continuous, seasonal mapping of $CH_4$, at levels of few parts per billion, at least in the brightest regions of the planet.

Having elaborated a complete pipeline for NOMAD spectra simulation is of great importance to extend the same analysis method to other trace species of interest, to serve as a solid background to plan current and forthcoming science operations with NOMAD, and to interpret science data integrating this forward model into optimal retrieval procedures.

## Acknowledgements

The NOMAD experiment is led by the Royal Belgian Institute for Space Aeronomy (IASB-BIRA), assisted by Co-PI teams from Spain (IAA-CSIC), Italy (INAF-IAPS), and the United Kingdom (Open University). This project acknowledges funding by the Belgian Science Policy Office (BELSPO), with the financial and contractual coordination by the ESA Prodex Office (PEA 4000103401, 4000121493), by Spanish MICIIN through its Plan Nacional (AYA2009-08190 and AYA2012-39691) and by European Funds under grant ESP2015-65064-C2-1-P (MINECO/FEDER), as well as by UK Space Agency through grants ST/R005761/1,

ST/P001262/1, ST/R001405/1 and ST/ R001405/1 and Italian Space Agency through grant 2018-2-HH.0.

This work was supported by NASA's Mars Program Office under WBS 604796, "Participation in the TGO/NOMAD Investigation of Trace Gases on Mars", and by NASA's SEEC initiative under Grant Number NNX17AH81A, "Remote sensing of Planetary Atmospheres in the Solar System and beyond".

*The NOMAD Team – Science Team*: Vandaele, Ann Carine; Lopez Moreno, Jose Juan; Bellucci, Giancarlo; Patel, Manish; Allen, Mark; Alonso-Rodrigo, Gustavo; Altieri, Francesca; Aoki, Shohei; Bauduin, Sophie; Bolsée, David; Clancy, Todd; Cloutis, Edward; Daerden, Frank; D'Aversa, Emiliano; Depiesse, Cédric; Erwin, Justin; Fedorova, Anna; Formisano, Vittorio; Funke, Bernd; Fussen, Didier; Garcia-Comas, Maia; Geminale, Anna; Gérard, Jean-Claude; Gillotay, Didier; Giuranna, Marco; Gonzalez-Galindo, Francisco; Hewson, Will; Homes, James; Ignatiev, Nicolai; Kaminski, Jacek; Karatekin, Ozgur; Kasaba, Yasumasa; Lanciano, Orietta; Lefèvre, Franck; Lewis, Stephen; López-Puertas, Manuel; López-Valverde, Miguel; Mahieux, Arnaud; Mason, Jon; Mc Connell, Jack; Mumma, Mike; Nakagawa, Hiromu, Neary, Lori; Neefs, Eddy; Novak, R.; Oliva, Fabrizio; Piccialli, Arianna; Renotte, Etienne; Robert, Severine; Sindoni, Giuseppe; Smith, Mike; Stiepen, Arnaud; Thomas, Ian; Trokhimovskiy, Alexander; Vander Auwera, Jean; Villanueva, Geronimo; Viscardy, Sébastien; Whiteway, Jim; Willame, Yannick; Wilquet, Valérie; Wolff, Michael; Wolkenberg, Paulina – *Tech Team*: Alonso-Rodrigo, Gustavo; Aparicio del Moral, Beatriz; Barzin, Pascal; Beeckman, Bram; BenMoussa, Ali; Berkenbosch, Sophie; Biondi, David; Bonnewijn, Sabrina; Candini, Gian Paolo; Clairquin, Roland; Cubas, Javier; Giordanengo, Boris; Gissot, Samuel; Gomez, Alejandro; Hathi, Brijen; Jeronimo Zafra, Jose; Leese, Mark; Maes, Jeroen; Mazy, Emmanuel; Mazzoli, Alexandra; Meseguer, Jose; Morales, Rafael; Orban, Anne; Pastor-Morales, M; Perez-grande, Isabel; Queirolo, Claudio; Ristic, Bojan; Rodriguez Gomez, Julio; Saggin, Bortolino; Samain, Valérie; Sanz Andres, Angel; Sanz, Rosario; Simar, Juan-Felipe; Thibert, Tanguy.